\documentclass[10pt, draftclsnofoot, twocolumn]{IEEEtran}
\pdfoutput=1
\usepackage{amssymb}
\usepackage{algorithm}
\usepackage{algpseudocode}
\usepackage{algorithmicx}
\usepackage{amsfonts}
\usepackage{amsmath}
\usepackage{amssymb}
\usepackage{amsthm}
\usepackage{array}
\usepackage{bbding}
\usepackage{bigints}
\usepackage{booktabs}
\usepackage{cite}
\usepackage{geometry}
\usepackage{color}
\usepackage{diagbox}
\usepackage{epsfig,latexsym}
\usepackage{epstopdf}
\usepackage{graphicx}
\usepackage{fancyhdr}
\usepackage{float}
\usepackage{indentfirst}
\usepackage{lastpage}
\usepackage{makecell}
\usepackage{mathtools}
\usepackage{multirow}
\usepackage{pifont}
\usepackage{psfrag}
\usepackage{setspace}
\usepackage{stfloats}
\usepackage[colorlinks,linkcolor=blue]{hyperref}
\hypersetup{hidelinks,
	colorlinks=true,
	allcolors=black,
	pdfstartview=Fit,
	breaklinks=true}
\usepackage{cleveref}
\renewcommand{\baselinestretch}{1.0}
\usepackage{empheq}
\usepackage{enumerate}
\usepackage{times}
\usepackage{subfigure}
\newcolumntype{C}{>{\centering\arraybackslash$}p{\linewidth}<{$}}
\newtheorem{theorem}{Theorem}

\newtheorem{lemma}{Lemma}

\newtheorem{corollary}{Corollary}

\DeclareMathOperator*{\argmin}{arg\,min}
\DeclareMathOperator{\diag}{diag}
\DeclareMathOperator{\tr}{tr}
\newcommand{\algorithmicbreak}{\textbf{break}}
\newcommand{\BREAK}{\State\algorithmicbreak}

\newcommand{\RNum}[1]{\uppercase\expandafter{\romannumeral #1\relax}}

\begin{document}
\title{Generalized Spatial Modulation Aided Affine Frequency Division Multiplexing}

% use a multiple-column layout for up to three different
\author{{Zeping Sui}, {\em Member,~IEEE}, {Zilong Liu}, {\em Senior Member,~IEEE}, {Leila Musavian}, {\em Member,~IEEE},\\ {Lie-Liang Yang}, {\em Fellow,~IEEE}, and {Lajos Hanzo}, {\em Life Fellow,~IEEE}
\vspace{-2em}
\thanks{Zeping Sui, Zilong Liu and Leila Musavian are with the School of Computer Science and Electronics Engineering, University of Essex, Colchester CO4 3SQ, U.K. (e-mail: zepingsui@outlook.com, \{zilong.liu,leila.musavian\}@essex.ac.uk).}% <-this % stops a space
\thanks{Lie-Liang Yang and Lajos Hanzo are with the Department of Electronics and Computer Science, University of Southampton, Southampton SO17 1BJ, U.K. (e-mail: lly@ecs.soton.ac.uk; lh@ecs.soton.ac.uk).}
}
\maketitle

\begin{abstract}
Generalized spatial modulation-aided affine frequency division multiplexing (GSM-AFDM) is conceived for reliable multiple-input multiple-output (MIMO) communications over doubly selective channels. \textcolor{black}{We commence by proposing several low-complexity detectors for large-scale GSM-AFDM systems to meet the diverse requirements of heterogeneous receiver designs in terms of detection complexity and reliability.} Specifically, we introduce the linear minimum mean square error (LMMSE) equalizer-based maximum likelihood detector (LMMSE-MLD). By exploiting the GSM properties, we then derive the LMMSE-based transmit-antenna activation pattern (TAP) check-based log-likelihood ratio detector (LMMSE-TC-LLRD). In addition, we propose a pair of new detectors, namely the greedy residual check detector (GRCD) and the reduced space check detector (RSCD). We also derive a bit error rate (BER) upper-bound by considering the MLD. \textcolor{black}{Our analytical results are also available for multiple-input multiple-output (MIMO)-AFDM, since MIMO-AFDM can be regarded as a specical case of the proposed GSM-AFDM.} Our simulation results demonstrate that 1) the BER upper bound derived is tight for moderate to high signal-to-noise ratios (SNRs), \textcolor{black}{2) the proposed GSM-AFDM achieves lower BER than its conventional orthogonal frequency division multiplexing (OFDM), orthogonal time frequency space (OTFS) and AFDM counterparts. Specifically, at a BER of $10^{-4}$ and a velocity of $540$ km/h, the proposed GSM-AFDM is capable of attaining about $6$ dB SNR gain compared to GSM-OFDM,} and 3) the conceived detectors strike a compelling trade-off between the BER and complexity.
\end{abstract}
%\vspace{-12mm}
%\vspace{-5mm}
\IEEEpeerreviewmaketitle

%%%%%%%%%%%%%%%%%%%%%%%%%%%%%%%%%%%%%%%%%%%%%%%%%%%%%%%%%
% Section1
\section{Introduction}\label{Section 1}
\subsection{\textcolor{black}{Background}}
The next-generation wireless systems are envisioned to support reliable communications over high-mobility channels \cite{9779322}. This involves a number of use cases, such as connected vehicles, integrated aerospace networks, etc. Compared to the fifth generation (5G) mobile systems, 6G is expected to support reliable information exchange even at the aircraft speed of $1000$ km/h \cite{Wang20236G}. At such a velocity, the legacy orthogonal frequency division multiplexing (OFDM) suffers from excessive inter-carrier interference (ICI) imposed by the Doppler phenomenon \cite{hanzo2005ofdm,9508932}.

To address the aforementioned problem, orthogonal time frequency space (OTFS) modulation has been proposed, as a benefit of its significantly improved error rate performance in high-mobility environments. The key idea of OTFS is to modulate data symbols in the delay-Doppler (DD)-domain and then convert them to the time-frequency (TF)-domain signal via the inverse symplectic finite Fourier transform (ISFFT) \cite{hadani2017orthogonal,8424569}. Thanks to the channel's extended coherence time and the OTFS signal properties, the DD-domain channel can be considered sparse and quasi-static during an OTFS frame \cite{10183832}. However, the OTFS transceiver complexity may be excessive, since it utilizes the two-dimensional orthogonal basis functions of ISFFT and SFFT \cite{9404861}. Moreover, the asymptotic diversity order of uncoded OTFS is one, as pointed out in \cite{8686339}. Hence, a strong channel code is preferred for high-performance OTFS transmission  \cite{9404861}.

\subsection{\textcolor{black}{Related Works}}
\subsubsection{\textcolor{black}{AFDM}} As an alternative to OTFS, affine frequency division multiplexing (AFDM) has been proposed in \cite{10087310,9473655}. In AFDM, the information bits are modulated in the discrete affine Fourier transform (DAFT)-domain, which may be considered a generalization of the discrete Fourier transform (DFT) of OFDM. From a TF-domain perspective, each information symbol is mapped to one of the orthogonal chirp carriers, traversing across the entire TF-domain. By carefully tuning the AFDM parameters according to the maximum Doppler, the non-zero elements of the effective channel matrices (associated with different paths) in the DAFT domain can avoid overlapping, and therefore the resultant system is capable of achieving full diversity \cite{10087310}. AFDM exhibits more convenient backward compatibility with OFDM, since DAFT can be efficiently implemented based upon the FFT. For accurate AFDM channel estimation, pilot chirps were embedded in the DAFT domain \cite{10087310}. A weighted maximal-ratio-combining based equalizer was also proposed in \cite{10087310} for exploiting the channel's diversity. A diagonal reconstruction-based channel estimation scheme was proposed in \cite{10557524}, whereby the DAFT-domain channel matrix can be directly estimated at an appealingly low complexity. In \cite{10711268}, a superimposed pilot scheme was invoked for AFDM channel estimation in order to enhance the system's spectral efficiency (SE). AFDM has also been integrated with index modulation \cite{10342712,10845819}. For example, the so-called cyclic delay diversity method was employed in \cite{10845819} for achieving a beneficial transmit diversity gain. \textcolor{black}{Moreover, upon the considering practical channel estimation schemes, machine learning-based channel estimation and data detection algorithms were developed in \cite{HUANG2025102597}.\footnote{\textcolor{black}{In general, the embedded channel estimation scheme proposed in \cite{10087310}, the embedded pilot-aided diagonal reconstruction channel estimation paradigm proposed in \cite{10557524}, and the machine learning-based channel estimation algorithm proposed in \cite{HUANG2025102597} can be leveraged in our proposed GSM-AFDM systems. Conceiving efficient channel estimation algorithms based on the sparse structure of the DAFT-domain channel matrices is an interesting topic, which motivates us to explore in our future work.}}} Motivated by the need for high-mobility machine-type communications, an AFDM-aided sparse code multiple access system was later proposed in \cite{10566604} for supporting massive connectivity over doubly selective channels. Based on the sparse structure of the DAFT-domain channel matrix, message-passing detectors were proposed in \cite{wu2024afdm}. \textcolor{black}{In \cite{10580928}, discrete Fourier transform (DFT)-based modulation and demodulation schemes were proposed for AFDM, whereby AFDM is interpreted as a precoded OFDM scheme. The BER performance of AFDM using minimum mean square error equalization (MMSE-Eq) was analyzed in \cite{10806672}, and two corresponding chirp parameter selection strategies were proposed to attain the optimal error rate performance. An algorithm for joint channel estimation, data detection (JCDE) and radar parameter estimation (RPE) was proposed for AFDM systems in \cite{10794592}, where a Bayesian parametric bilinear Gaussian belief propagation framework was developed to perform JCDE, while probabilistic data association (PDA) and Bernoulli-Gaussian denoising were combined to carry out RPE. More recently, an AFDM-based integrated sensing and communication system was proposed in \cite{10858612}, where two new performance metrics, including sensing spectral efficiency (SSE) and sensing outage probability (SOP), were conceived. Then the relationship between the above two metrics and AFDM waveform parameters was derived, and a trade-off between sensing and communication performance was observed. Moreover, a novel matched filter-based delay and Doppler estimation method was proposed. Later in \cite{10975107}, pre-chirp index modulation was combined with AFDM systems, yielding improved BER performance. Specifically, the average bit error probability and upper bounds were derived, and an optimal pre-chirp alphabet was conceived for improving the BER performance. The random access preamble transmission based on AFDM for mobile satellite communication systems was conceived in \cite{10996536}, where an algorithm was proposed for enhancing the timing detection accuracy and it was illustrated that AFDM can eliminate the carrier frequency offset of Zadoff-Chu sequences.}
%%%%%%%%%%%%%%%%%%%%%%%%%%%%%%%%%%%%%%%%%%%%%%%
% table 1
\begin{table*}[t]
%\vspace{-3mm}
\footnotesize
\begin{center}
\caption{\textcolor{black}{Contrasting our contributions to the related AFDM literature}}
\label{table1}
\begin{tabular}{l|c|c|c|c|c|c|c|c|c|c|c}
\hline
Contributions & \textbf{This paper} & \cite{10087310} & \cite{9473655} & \cite{10557524} & \cite{10342712} & \cite{10845819} & \cite{10566604} & \cite{wu2024afdm} & \cite{10806672} & \cite{10794592} & \cite{10975107} \\
\hline
\hline
GSM-AFDM & \checkmark & &   &  &  &  & & & & \\  
\hline
BER performance analysis & \checkmark & \checkmark & \checkmark & \checkmark & \checkmark & \checkmark & \checkmark & \checkmark & \checkmark &  & \checkmark  \\  
\hline
Diversity order analysis & \checkmark & \checkmark & \checkmark & \checkmark & & \checkmark & \checkmark &  & \checkmark & & \checkmark \\  
\hline
Coding gain analysis & \checkmark & & & & & \checkmark & \checkmark & & &  \\  
\hline 
Capacity analysis & \checkmark & & & & & & & & & \\  
\hline
Statistics of transmitted symbols & \checkmark &  &  & & & \checkmark & & \checkmark & \checkmark & \checkmark & \checkmark \\ 
\hline
Detector complexity analysis & \checkmark & \checkmark & & &  & \checkmark & \checkmark & \checkmark & \checkmark & \checkmark & \\ 
\hline
LDPC coded system & \checkmark & & & & \checkmark & & \checkmark & & & \\
\hline 
\textbf{TAP checking} & \checkmark & & & & & & & & & \\ 
\hline
\textbf{LLR-based detector} & \checkmark & & & & & & & & & \\ 
\hline
\textbf{Greedy residual check detector} & \checkmark & & & & & & & & & \\ 
\hline
\textbf{Reduced space check detector} & \checkmark & & & & & & & & & \\ 
\hline
\end{tabular}
\end{center}
\vspace{-4em}
\end{table*}
%%%%%%%%%%%%%%%%%%%%%%%%%%%%%%%%%%%%%%%%%%%%%%%%%%%%%%%%%

\subsubsection{\textcolor{black}{SM and GSM}} As a parallel development, spatial modulation (SM) constitutes an attractive paradigm for multiple-input multiple-output (MIMO) communications \cite{4382913,6678765}. In principle, only a single transmit antenna (TA) is activated over each SM time-slot. In addition to the classic amplitude-phase modulated (APM) symbols, extra information bits are mapped onto the specific index of the instantaneously activated TA. Therefore, SM systems are capable of avoiding inter-antenna interference and achieving high energy efficiency (EE), while the tight requirement of inter-antenna synchronization can be relaxed \cite{7458894}. Very recently, SM has also been applied to reconfigurable intelligent surface (RIS)-aided systems \cite{10640072}, where only part of the TAs or RIS elements are activated. In this way, a flexible SE vs. EE trade-off may be attained using the reflection modulation technique concept of \cite{10640072}. To further improve the data rate of SM systems, generalized SM (GSM) has been studied in \cite{5757786,6166339}, in which a few (at least one) TAs are activated to transmit APM symbols, and the remaining bits can be mapped onto the TA activation patterns (TAPs). \textcolor{black}{Upon leveraging polarization dimension at the transmitter and polarized receive antennas, the GSM system can obtain polarization diversity, yielding improved channel capacity \cite{9398857}.} The SM-aided OTFS systems were investigated in \cite{10129061,9521176,10250854}, although GSM was not considered. However, neither the inter-symbol interference (ISI) nor the ICI introduced by high-mobility channels was fully considered in \cite{9521176}. Later in \cite{10531245}, TA selection schemes were combined with SM-OTFS systems for achieving transmit diversity and for improving \textcolor{black}{the bit error ratio (BER) performance}.

\subsection{\textcolor{black}{Motivations \& Contributions}}
\textcolor{black}{We explore the synergistic integration of GSM and AFDM, termed as GSM-AFDM, for supporting reliable MIMO communications over high-mobility channels. Our research is driven by the following fundamental questions:
\begin{itemize}
	\item Can the BER performance of AFDM systems be further improved by harnessing the SM/GSM techniques, since SM/GSM-aided systems are capable of improving the error rate performance of conventional MIMOs \cite{7063946}? If so, how to analyze it quantitatively?
	\item Conventional OFDM based GSM systems suffer from significant performance loss in high-mobility scenarios, since OFDM is sensitive to the ICI introduced by Doppler effect. Can we exploit the Doppler resilience of AFDM to further improve the error rate performance of GSM systems over doubly-selective fading channels?
	\item In view of the diverse quality-of-service (QoS) requirements for different types of services in modern communication systems, how to develop a variety of GSM-AFDM detectors to strike an attractive trade-off between detection complexity and receiver reliability?
	\end{itemize}}

\textcolor{black}{In contrast to the related papers in Table \ref{table1}, the main contributions of this work are summarized as follows:}
\textcolor{black}{\begin{itemize}
	\item We conceive the intrinsically amalgamated GSM-AFDM concept for reliable MIMO communications in high-Doppler scenarios, where the information bits are conveyed by both the classic APM symbols and the TAP-indices. Based on the proposed GSM symbol mapper, the DAFT-domain symbols can be mapped to a dedicated sparse signal frame.
	\item We then propose several low-complexity detectors for large-scale GSM-AFDM systems. Firstly, by invoking a bespoke linear minimum mean square error (LMMSE) equalizer, we derive both the LMMSE-based MLD and the LMMSE-based TAP checking log-likelihood ratio detector (LMMSE-TC-LLRD) for each group, where the concept of TAP checking (TC) is proposed for avoiding catastrophic TAP decisions. Moreover, inspired by the philosophy of greedy algorithms, the greedy residual check detector (GRCD) is proposed, where the reliabilities of chirp subcarriers are checked and we can avoid repeated TAP checkings. Furthermore, upon checking the reliabilities of all the TAPs, we propose a reduced-space check detector (RSCD), where only a subset of TAPs have be tested.
	\item We derive a closed-form BER upper-bound expression using the union-bound technique and the MLD. We demonstrate that the upper-bound is tight in the moderate to high SNR regions. Since MIMO-AFDM can be considered as a special case of GSM-AFDM, our analytical results in Section IV are also applicable to MIMO-AFDM systems. Afterwards, we analyze the diversity order and effective coding gain of the proposed GSM-AFDM systems. We then derive the discrete-input continuous-output memoryless channel (DCMC) capacity of our GSM-AFDM.
	\item Our simulation results illustrate the superiority of our proposed GSM-AFDM systems and low-complexity detectors in that 1) the proposed GSM-AFDM scheme attains improved BER performance and DCMC capacity compared to its conventional SM- and SIMO-based AFDM counterparts; 2) the proposed TC technique beneficially improves the BER performance, and all the proposed detectors can attain near-LMMSE-MLD BER performance and strike flexible trade-offs between BER and complexity; 3) the proposed GSM-AFDM outperforms both GSM-OFDM and GSM-OTFS, thanks to its higher diversity order, indicating the advantage of GSM-AFDM under high-mobility scenarios.
\end{itemize}}

The rest of our paper is organized as follows. The system model of our GSM-AFDM is investigated in Section \ref{Section 2}. Our low-complexity detectors and the corresponding complexity analysis are detailed in Section \ref{Section 3}. In Section \ref{Section 4}, we characterize the overall system performance, while our simulation results are offered in Section \ref{Section 5}. Finally, our conclusions are formulated in Section \ref{Section 6}.

\emph{Notation:} $\mathbb{Z}_{+}^{M_t}$ indicates the real integer set of $\{1,\ldots,M_t\}$. Lower- and Upper-case boldface letters represent \textcolor{black}{vectors and matrices}, respectively; the complex Gaussian distribution with mean vector $\pmb{a}$ and covariance matrix $\pmb{B}$ is denoted as $\mathcal{CN}(\pmb{a},\pmb{B})$. Moreover, $\mathbb{B}$ stands for the bit set of $\left\{0,1\right\}$ and $\lfloor\cdot\rfloor$ is the flooring operator; $\Re\left\{\cdot\right\}$ and $[\cdot]_{N}$ indicate taking the real part and the modulo-$N$ operator, respectively; the uniform distribution within the interval $[a,b]$ is represented by $\mathcal{U}[a,b]$. Furthermore, the Kronecker product operator is denoted by $\otimes$; $(\pmb{A})^T$, $(\pmb{A})^H$ and $(\pmb{A})^{-1}$ are the transpose, conjugate transpose and inverse of the matrix $\pmb{A}$, respectively. \textcolor{black}{$||\cdot||$ stands the vector Euclidean norm and $\odot$ is the Hadamard product.} Finally, $\pmb{I}_N$ and $\pmb{e}_N(n)$ denote the $N$-dimensional identity matrix and its $n$th column, respectively, while $\delta(\cdot)$ is the delta function; $\binom{M_t}{K}$ represents the possible number of combinations by choosing $K$ out of $M_t$.
%%%%%%%%%%%%%%%%%%%%%%%%%%%%%%%%%%%%%%%%%%%%%%%%%%%%%%%%%
% Section2
\section{System Model}\label{Section 2}
\subsection{Transmitter Signals}
\begin{figure*}[t]
\centering
\includegraphics[width=0.9\linewidth]{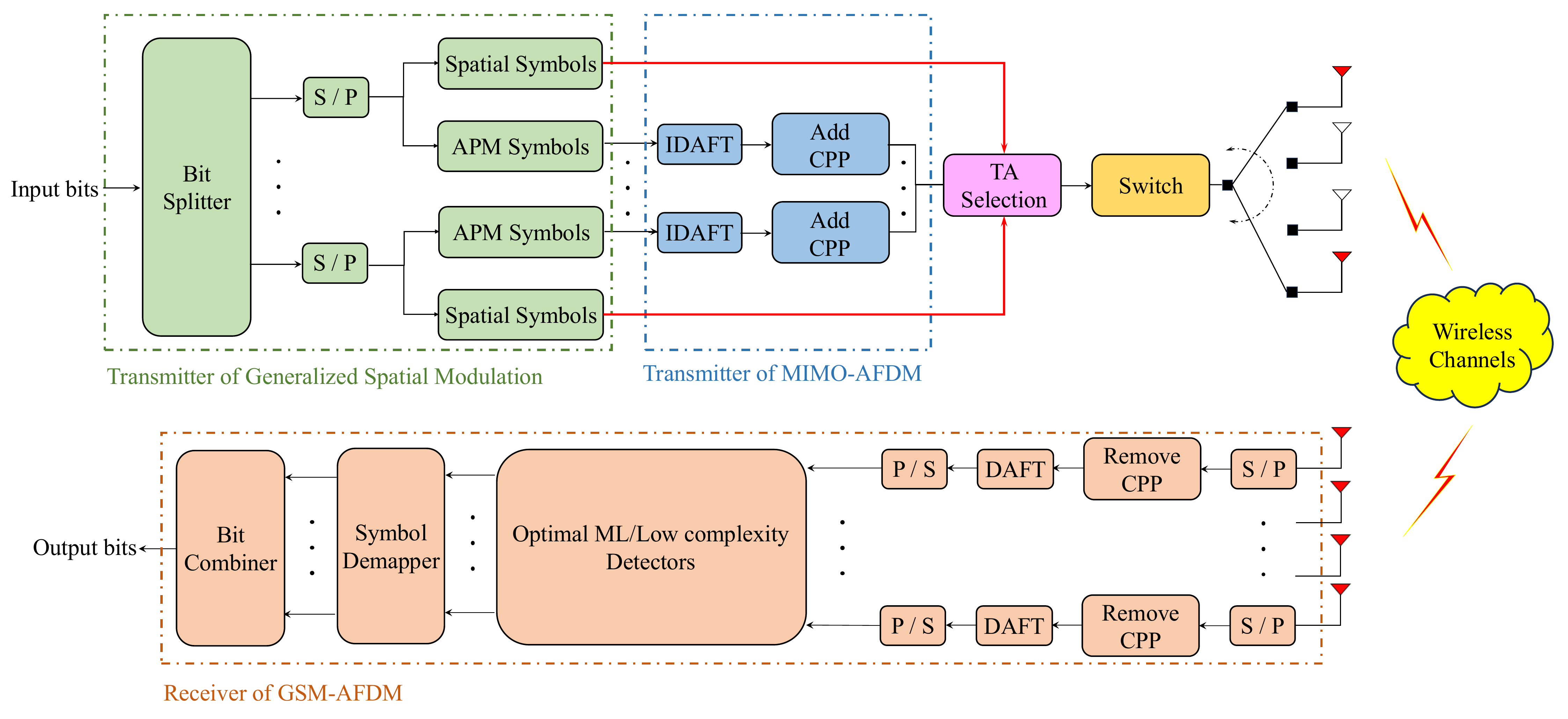}
\vspace{-1em}
\caption{\textcolor{black}{Illustration of a toy example of the GSM-AFDM system, where $K=2$ out of $M_t=4$ TAs and $M_r=4$ RAs are activated.}}
\label{Figure1}
\vspace{-2em}
\end{figure*}

As demonstrated in Fig. \ref{Figure1}, we consider a MIMO-AFDM system including $M_t$ TAs and $M_r$ receive antennas (RAs). Specifically, we consider an AFDM system, which has $N$ chirp subcarriers with a subcarrier spacing $\Delta f$ and the symbol duration $T=1/\Delta f$. Consequently, the sampling interval, i.e., delay resolution, can be formulated as $T_s=1/B=1/(N\Delta f)$, given the bandwidth of $B=N\Delta f$ and the sampling rate of $f_s=1/T_s$. We assume that an $L$-length bit sequence $\pmb{b}\in\mathbb{B}^L$ is transmitted which is divided into $N$ groups, yielding $L_b=L/N=L_1+L_2$ in each group. By denoting the bit sequence as $\pmb{b}=[\pmb{b}_1,\ldots,\pmb{b}_N]$, we have the $n$th component $\pmb{b}_n=[\pmb{b}_{n,1},\pmb{b}_{n,2}]$. In each group, we assume that $K$ out of $M_t$ TAs are activated, yielding $L_1\leq\left\lfloor\log_2 \binom{M_t}{K}  \right\rfloor$. Consequently, there are $C=2^{L_1}$ possible TAPs in the $n$th group, which can be written as $\mathcal{X}=\{\mathcal{X}_1,\ldots,\mathcal{X}_C\}$. Let the $c$th TAP be $\mathcal{X}_c=\{\mathcal{X}_{c,0},\ldots,\mathcal{X}_{c,(K-1)}\}$, where $\mathcal{X}_{c,k}\in\mathbb{Z}^{M_t}_{+}$ for $k=0,\ldots,K-1$. Then, the TAP invoked can be formulated as $\mathcal{I}_n=\mathcal{X}_c\subset\mathcal{X}$. The remaining bit sequence $\pmb{b}_{n,2}\in\mathbb{B}^{L_2}$ associated with $L_2=K\log_2 Q$ is mapped onto $K$ classic APM symbols, based on the $Q$-ary normalized constellation $\mathcal{A}=\{a_1,\ldots,a_Q\}$. Therefore, the attainable rate can be formulated as
\begin{align}
	L_b=\left\lfloor\log_2\binom{M_t}{K}\right\rfloor+K\log_2 Q\ \text{bits/s/subcarrier}.
\end{align}
Then, the pure-data symbol vector of group $n$ can be expressed as $\pmb{x}_n^d=[x_n^d(0),\ldots,x_n^d(K-1)]^T$ along with \textcolor{black}{$\mathbb{E}[|x_{n}^d(k)|^2]=1$} and $\forall x_{d,n}(k)\in\mathcal{A}$, for $k=0,\ldots,K-1$. According to the TAP $\mathcal{I}_n$, the symbols in $\pmb{x}_n^d$ are assigned to an $M_t$-length transmit vector $\pmb{x}_n$, which can be formulated as $\pmb{x}_n=\pmb{\Upsilon}_{\mathcal{I}_n}\pmb{x}_n^d$ with the $(M_t\times K)$-dimensional mapping matrix $\pmb{\Upsilon}_{\mathcal{I}_n}$ based on $\mathcal{I}_n$. Following from the above analysis, the bit-to-symbol mapping relationship can be represented by a codebook, yielding
\begin{align}\label{eq_codebook}
	\mathcal{D}\triangleq\{\pmb{d}_1,\ldots,\pmb{d}_{2^{L_b}}:\pmb{d}_i\in\mathbb{C}^{M_t},i=1,\ldots,2^{L_b}\}.
\end{align}
%%%%%%%%%%%%%%%%%%%%%%%%%%%% Figure 1 %%%%%%%%%%%%%%%%%%%%%%%%%%%%%%%
%%%%%%%%%%%%%%%%%%%%%%%%%%%%%%%%%%%%%%%%%%%%%%%%%%%%%%%%%%%%%%%%%%%%%
The $(M_t\times N)$-dimensional DAFT-domain AFDM frame $\pmb{X}=[\pmb{x}_0,\ldots,\pmb{x}_{N-1}]$ can be expressed as
\begin{align}\label{eq_X}
	\pmb{X}=\begin{bmatrix}
		X(0,0) & \cdots & X(0,N-1) \\
		\vdots & \ddots & \vdots \\
		X(M_t-1,0) & \cdots & X(M_t-1,N-1)
	\end{bmatrix},
\end{align}
where the $m_t$th row ${\pmb{X}}_{m_t,:}\in\mathbb{C}^{1\times N}$ is transmitted by the $m_t$th TA for $m_t=0,\ldots,M_t-1$, and each column only has $K$ non-zero elements.

\textcolor{black}{\emph{Example 1:} Consider a GSM-AFDM system with $N=4$ subcarriers, where $K=2$ out of $M_t=4$ TAs are activated and using QPSK modulation, yielding $L_1=2$ bits and $L_2=4$. Suppose we have the overall bit sequence as $\pmb{b}=[000010101001010001110111]$. Specifically, we have $\pmb{b}_{1,1}=[00]$ and $\pmb{b}_{1,2}=[0010]$ for the first group, then the first column of $\pmb{X}$ can be expressed as $\pmb{x}_0=[-\sqrt{2}/2+\sqrt{2}/2j,\sqrt{2}/2+\sqrt{2}/2j,0,0]^T$. Then, the entire AFDM frame $\pmb{X}=[\pmb{x}_0,\pmb{x}_1,\pmb{x}_0,\pmb{x}_3]$ can be given by
\begin{align}
	\begin{bmatrix}
		-\frac{\sqrt{2}}{2}+\frac{\sqrt{2}}{2}j & \frac{\sqrt{2}}{2}+\frac{\sqrt{2}}{2}j & 0 & 0\\
		\frac{\sqrt{2}}{2}+\frac{\sqrt{2}}{2}j & 0 & \frac{\sqrt{2}}{2}+\frac{\sqrt{2}}{2}j & 0\\
		0 & -\frac{\sqrt{2}}{2}-\frac{\sqrt{2}}{2}j & 0 & -\frac{\sqrt{2}}{2}-\frac{\sqrt{2}}{2}j\\
		0 & 0 & 	-\frac{\sqrt{2}}{2}-\frac{\sqrt{2}}{2}j & \frac{\sqrt{2}}{2}-\frac{\sqrt{2}}{2}j			\end{bmatrix}.\nonumber
\end{align}}

Upon utilizing IDAFT, and letting ${\pmb{z}}_{m_t}={\pmb{X}}_{m_t,:}^T$, the time-domain (TD) signal transmitted from the $m_t$th TA can be formulated based on the IDAFT as
\begin{align}\label{eq_s}
	s_{m_t}(n)=\sum_{q=0}^{N-1}z_{m_t}(q)\phi_n(q),\quad n=0,\ldots,N-1,
\end{align}
where $\phi_n(q)=e^{j2\pi(c_1 n^2+c_2 q^2+nq/N)}/\sqrt{N}$ represents the transform kernel corresponding to the $q$th chirp subcarrier and the $n$th TD symbol having the AFDM parameters $c_1$ and $c_2$. Moreover, \eqref{eq_s} can be rewritten as
\begin{align}\label{eq_s_vector}
	\pmb{s}_{m_t}=\pmb{A}^H{\pmb{z}}_{m_t}=\pmb{\Lambda}^H_{c_1}\pmb{\mathcal{F}}^H\pmb{\Lambda}^H_{c_2}{\pmb{z}}_{m_t},
\end{align}
where $\pmb{A}=\pmb{\Lambda}_{c_2}\pmb{\mathcal{F}}\pmb{\Lambda}_{c_1}$ denotes the DAFT matrix associated with $\pmb{\Lambda}_{c}=\diag\left(1,e^{-j2\pi c},\ldots,e^{-j2\pi c(N-1)^2}\right)$ for the DAFT parameter $c\in\{c_1,c_2\}$. To mitigate the inter-symbol interference and invoke circulant convolution, an $L_P$-length chirp-periodic prefix (CPP) is employed within the AFDM frame, yielding
\begin{align}
	s_{m_t}(n)=s(N+n)e^{-j2\pi c_1(N^2+2Nn)},
	\end{align}
for $n=-L_{P},\ldots,-1$. Consequently, we can obtain the $m_t$th TD transmit signal $s_{m_t}(t)$.
%\vspace{-3em}
\subsection{Channel Description}
We consider a doubly-selective channel having $P$ paths, whose delay-time (DT)-domain channel impulse response spanning from the TA $m_t$ to the RA $m_r$ can be expressed as
\begin{align}\label{eq_h}
	h_{m_r,m_t}(t,\tau)=\sum_{p=1}^P h_{p,m_r,m_t}e^{-j2\pi\nu_p t}\delta(\tau-\tau_p),
\end{align}
where $h_{p,m_r,m_t}$, $\nu_p$ and $\tau_p$ denote the channel gain, delay shifts and Doppler shifts of the $p$th path, respectively. Consequently, the $p$th Doppler and delay indices can be respectively formulated as
\begin{align}
	\nu_p=\frac{k_p}{T}=\frac{k_pf_s}{N},\ \tau_p=\frac{l_p}{N\Delta f}={l_p}{T_s},
\end{align}
where $f_s=N \Delta f$ is the sampling rate and $T_s=1/f_s=T/N$ denotes the sampling interval. Explicitly, the normalized Doppler and delay shifts can be respectively denoted by $k_p=\alpha_p+\beta_p$ and $l_p$, where the integer components are $\alpha_p\in[-\alpha_{\text{max}},\alpha_{\text{max}}]$ and $l_p\in[0,l_{\text{max}}]$, while $\beta_p\in[-1/2,1/2]$ represents the fractional Doppler shift. Moreover, we have $l_\text{max}=\max(l_i)=L_P<N$ and the maximum delay $\tau_\text{max}={l_\text{max}}/{(N\Delta f)}$. The discrete DT-domain channel response can be obtained by sampling at $t=nT_s$ for $0\leq n\leq N-1$, yielding
\begin{align}\label{eq_h_dis}
	h_{m_r,m_t}(n,q)=\sum_{p=1}^P h_{p,m_r,m_t}e^{-j2\pi\frac{\nu_p}{f_s}n}\delta\left(q-\frac{\tau_p}{T_s}\right),
\end{align}
where $q=\tau/T_s$ represents the normalized delay index.
\vspace{-1em}
\subsection{Received Signals}
The $m_r$th continuous TD signal received from the $m_t$th TA can be formulated based on \eqref{eq_h} as
\begin{align}\label{eq_TD_r}
	r_{m_r,m_t}(t)=\int_0^{\tau_\text{max}}h_{m_r,m_t}(t,\tau)s_{m_t}(t-\tau)d\tau+\bar{w}_{m_r,m_t}(t),
\end{align}
where $\bar{\omega}(t)$ denotes the complex additive white Gaussian noise (AWGN) term in the TD. Based on \eqref{eq_h_dis}, after sampling at $\{t=nT_s,n=0,\ldots,N-1\}$ and discarding the CPP, the discrete TD signal received at the $m_r$th RA from the $m_t$th TA can be expressed as
\begin{align}\label{eq_r_element}
	r_{m_r,m_t}(n)=\sum_{q=0}^{\infty}s_{m_t}(n-q)h_{m_r,m_t}(n,q)+\bar{w}_{m_r,m_t}(n).
\end{align}
Therefore, the $m_r$th received signal corresponding to the $m_t$th TA can be written in matrix form as
\begin{align}\label{eq_r_vector}
	\pmb{r}_{m_r,m_t}=\bar{\pmb{H}}_{m_r,m_t}\pmb{s}_{m_t}+\bar{\pmb{w}}_{m_r,m_t},
\end{align}
where $\bar{\pmb{w}}_{m_r}$ denotes the corresponding AWGN vector. The TD channel matrix can be formulated as \cite{10087310}
\begin{align}\label{eq_H_time}
	\bar{\pmb{H}}_{m_r,m_t}=\sum_{p=1}^P h_{p,m_r,m_t}\pmb{\Gamma}_{\text{CPP}_p}\pmb{\Delta}_{k_p}\pmb{\Pi}^{l_p},
\end{align}
where $\pmb{\Pi}$ denotes the permutation matrix associated with forward cyclic shift, yielding
\begin{align}\label{eq_pi}
	\pmb{\Pi}=\begin{bmatrix}
		0 & \cdots & 0 & 1 \\
		1 & \cdots & 0 & 0 \\
		\vdots & \ddots & \ddots & \vdots \\
		0 & \cdots & 1 & 0		
	\end{bmatrix}_{N\times N},
\end{align}
while $\pmb{\Delta}_{k_p}=\diag\left\{1,e^{-j2\pi k_p/N},\ldots,e^{-j2\pi k_p(N-1)/N}\right\}$ is defined to characterize the Doppler effect and $\pmb{\Gamma}_{\text{CPP}_p}=\diag\{\rho_0,\ldots,\rho_N\}$ with
\begin{align}\label{eq_Gamma_element}
	\rho_n=\begin{cases}
		e^{-j2\pi c_1 [N^2-2N(l_p-n)]}, & n<l_p,\\
		1, & n\geq l_p,
			\end{cases}
\end{align}
for $n=0,\ldots,N-1$. By exploiting DAFT, the $m_r$th signal received in the DAFT-domain from the $m_t$th TA can be presented as
\begin{align}\label{eq_SISO}
	\pmb{y}_{m_r,m_t}=\pmb{A}\pmb{r}_{m_r,m_t}=\pmb{H}_{m_r,m_t}\pmb{z}_{m_t}+\pmb{w}_{m_r,m_t},
\end{align}
where $\pmb{H}_{m_r,m_t}=\pmb{A}\bar{\pmb{H}}_{m_r,m_t}\pmb{A}^H$ is the DAFT-domain channel matrix, which will be discussed later. The signal received at the $m_r$th RA is given by $\pmb{y}_{m_r}=\sum_{m_t =0}^{M_t-1}\pmb{y}_{m_r,m_t}=\sum_{m_t =0}^{M_t-1}\pmb{H}_{m_r,m_t}{\pmb{z}}_{m_t}+\pmb{w}_{m_r}$, where $\pmb{w}_{m_r}\sim\mathcal{CN}(\pmb{0},N_0\pmb{I}_N)$ is the DAFT-domain AWGN vector with $N_0=K/(\gamma_sM_t)$, where $\gamma_s$ denotes the SNR per symbol. The DAFT-domain channel matrix can be expressed based on \eqref{eq_H_time} as $\pmb{H}_{m_r,m_t}=\sum_{p=1}^Ph_{p,m_r,m_t}\pmb{H}_p$, where $\pmb{H}_p=\pmb{A}\pmb{\Gamma}_{\text{CPP}_p}\pmb{\Delta}_{k_p}\pmb{\Pi}^{l_p}\pmb{A}^H$. Upon substituting \eqref{eq_s}, \eqref{eq_h} and \eqref{eq_r_element} into \eqref{eq_SISO}, the element-wise DAFT-domain input-output relationship can be formulated as
\textcolor{black}{\begin{align}\label{eq_y_element}
	y_{m_r,m_t}(a)=&\sum_{p=1}^P\sum_{b=0}^{N-1}H_p(a,b)x_{m_t}(b)+w_{m_r}(a)\nonumber\\
	=&\frac{1}{N}\sum_{p=1}^P\sum_{b=0}^{N-1}h_{p,m_r,m_t}\eta(l_p,a,b)\nonumber\\
	&\times\zeta(l_p,k_p,a,b)x_{m_t}(b)+w_{m_r}(a),
\end{align}}
\hspace{-0.5em}where we have $\eta(l_p,a,b)=e^{j\frac{2\pi}{N}\left[N c_1 l_p^2-bl_p+Nc_2(b^2-a^2)\right]}$, and the spreading factor introduced by fractional Doppler shifts can be formulated as
\begin{align}
	\zeta(l_p,k_p,a,b)=\frac{e^{-j2\pi(a-b+\text{Ind}_p)}-1}{e^{-j\frac{2\pi}{N}(a-b+\text{Ind}_p)}-1},
\end{align}
and having the index indicator $\text{Ind}_p=(k_p+2Nc_1 l_p)_N$. Explicitly, there are only \textcolor{black}{$k_\text{non}=2k_{\nu}+1$} non-zero elements in each column or row of $\pmb{H}_p$ along with the fractional Doppler indicator parameter $k_{\nu}$. Furthermore, the central point indices of non-zero elements are given by $a^\text{peak}_p=\text{round}[(a+\text{Ind}_p)_N]$ \cite{10087310}. Hence, the DAFT-domain input-output relationship of \eqref{eq_y_element} can be rewritten as \cite{10087310}
\begin{align}
	y_{m_r,m_t}(a)&\approx\frac{1}{N}\sum_{p=1}^P\sum_{b=(a^\text{peak}_p-k_\nu)_N}^{(a^\text{peak}_p+k_\nu)_N}h_{p,m_r,m_t}\eta(l_p,a,b)\nonumber\\
	&\times\zeta(l_p,k_p,a,b)x_{m_t}(b)+w_{m_r}(a).
\end{align}
To achieve full diversity, the parameter $c_1$ should satisfy
\begin{align}\label{eq_c1}
	c_1=\frac{2(\alpha_\text{max}+k_\nu)+1}{2N},
\end{align}
where the parameter $k_\nu$ satisfies
\begin{align}\label{eq_N}
	2(\alpha_\text{max}+k_\nu)(l_\text{max}+1)+l_\text{max}<N,
\end{align}
%%%%%%%%%%%%%%%%%%%%%%%%%%%% Figure 2 %%%%%%%%%%%%%%%%%%%%%%%%%%%%%%%
\begin{figure}[t]
%\vspace{-1em}
\centering
\includegraphics[width=0.55\linewidth]{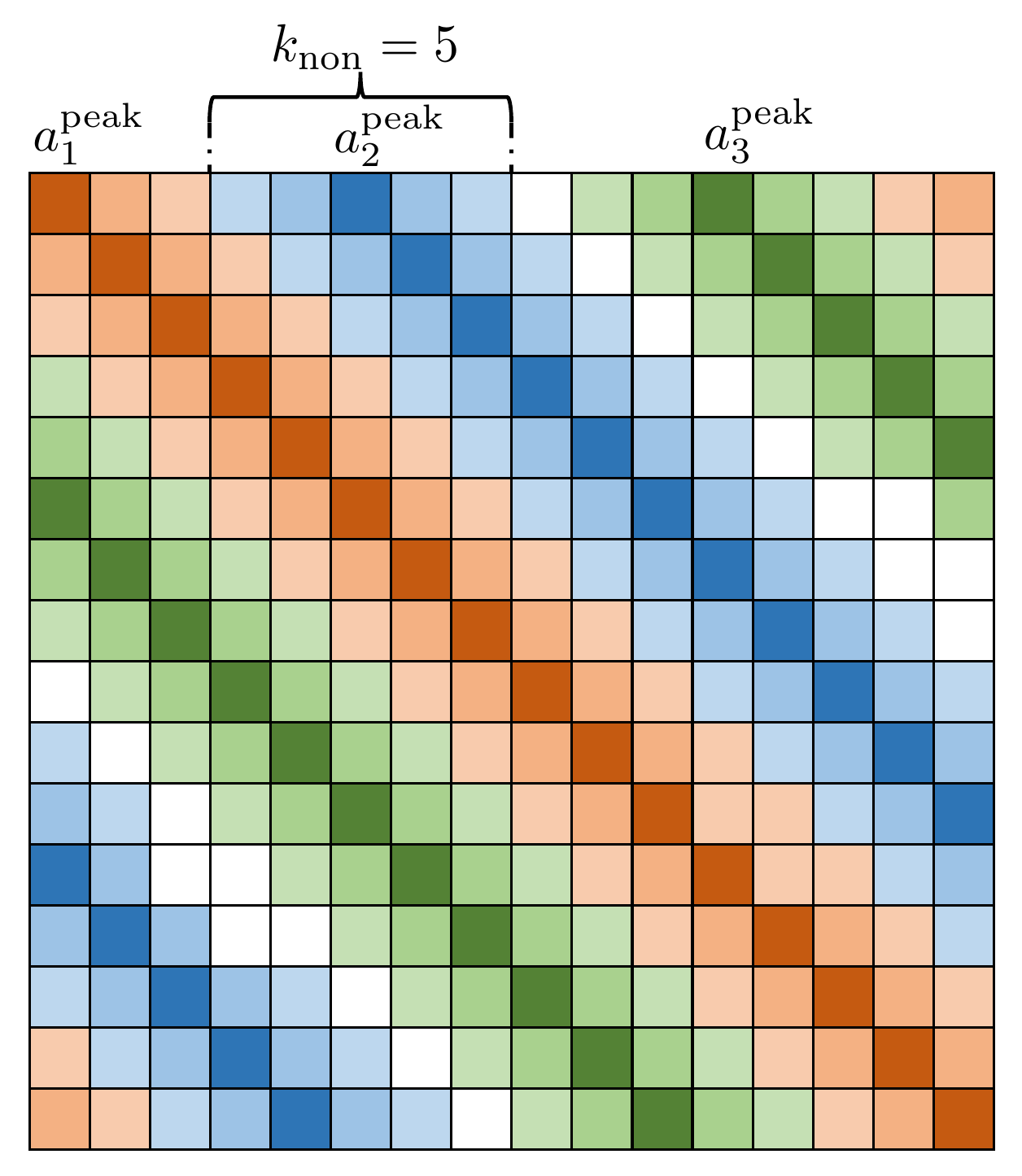}
%\vspace{-1em}
\caption{The DAFT-domain channel matrix $\pmb{H}_{m_r,m_t}$ with $N=16$ and \textcolor{black}{$k_{\nu}=1$, yielding $k_\text{non}=2k_\nu+1=5$.} Moreover, we use $l_\text{max}=2$, $\alpha_\text{max}=1$, $\mathcal{K}=\{k_1,k_2,k_3\}=\{0.2,0.3,1.4\}$ and $\mathcal{L}=\{l_1,l_2,l_3\}=\{0,1,2\}$.}
\label{Figure2}
%\vspace{-1em}
\end{figure}
%%%%%%%%%%%%%%%%%%%%%%%%%%%%%%%%%%%%%%%%%%%%%%%%%%%%%%%%%%%%%%%%%%%%%
\hspace{-0.4em}and we have $k_{\nu}=0$ in purely integer Doppler shift scenarios. Furthermore, $c_2$ can be set as an arbitrary irrational or a rational number sufficiently smaller than $1/(2N)$. Under the above $c_1$ and $c_2$ scenarios, there is no overlap between the non-zero elements of different $\pmb{H}_p$ within $\pmb{H}_{m_r,m_t}$. \textcolor{black}{Specifically, in Fig. \ref{Figure2}, the sub-blocks having different colors correspond to different matrices $\pmb{H}_p$ for $p=1,\ldots,P$.} We denote the DAFT-domain stacked transmit symbol vector and the stacked noise vector respectively as ${\pmb{z}}=[{\pmb{z}}_0^T,\ldots,{\pmb{z}}_{M_t-1}^T]^T\in\mathbb{C}^{NM_t}$ and $\pmb{w}=[\pmb{w}_0^T,\ldots,\pmb{w}_{M_r-1}^T]^T\in\mathbb{C}^{NM_r}$. The DAFT-domain MIMO channel matrix $\pmb{H}\in\mathbb{C}^{NM_r\times NM_t}$ is given by
\begin{align}\label{eq_MIMO_channel}
	\pmb{H}=\begin{bmatrix}
		\pmb{H}_{0,0} & \cdots & \pmb{H}_{0,M_t-1}\\
		\vdots & \ddots & \vdots\\
		\pmb{H}_{M_r-1,0} & \cdots & \pmb{H}_{M_r-1,M_t-1}			
		\end{bmatrix}.
\end{align}
Explicitly, when the AFDM system can attain full diversity, each row and column of $\pmb{H}_{m_r,m_t}$ only has $P(2k_\nu +1)$ non-zero elements. Consequently, the DAFT-domain end-to-end input-output relationship can be formulated as
\begin{align}\label{eq_y}
	\pmb{y}=\pmb{H}{\pmb{z}}+\pmb{w},
\end{align}
where $\pmb{y}=[\pmb{y}_0^T,\ldots,\pmb{y}_{M_r-1}^T]^T\in\mathbb{C}^{NM_r}$ denotes the DAFT-domain received stacked vector. To exploit the characteristic of GSM, we introduce the GSM permutation matrix ${\pmb{P}}\in\mathbb{C}^{NM_t\times NM_t}$, which can be formulated as
\begin{align}\label{eq_shuffled}
	{\pmb{P}}=\begin{bmatrix}
		\pmb{I}_{M_t}\otimes\pmb{e}^T_{N}(0)\\
		\vdots\\
		\pmb{I}_{M_t}\otimes\pmb{e}^T_{N}(N-1)
			\end{bmatrix}^T.
\end{align}
Therefore, we have ${\pmb{z}}=\pmb{P}{\pmb{x}}$ with $\pmb{x}=[\pmb{x}_0^T,\ldots,\pmb{x}_{N-1}^T]^T$, and \eqref{eq_y} can be rewritten as
\begin{align}\label{eq_y_re}
	\pmb{y}=\pmb{H}\pmb{P}\pmb{x}+\pmb{w}=\pmb{G}\pmb{x}+\pmb{w},
\end{align}
where $\pmb{G}=\pmb{H}\pmb{P}$ denotes the effective DAFT-domain channel matrix. Consequently, the conditional probability density function (PDF) of $\pmb{y}$ given $\pmb{x}$ can be formulated based on \eqref{eq_y} as
\begin{align}\label{eq_pdf}
	p(\pmb{y}|\pmb{x})=\frac{1}{(\pi N_0)^{NM_r}}\exp\left(-\frac{||\pmb{y}-\pmb{G}\pmb{x}||^2}{N_0}\right).
\end{align}
\section{Signal Detection in GSM-AFDM Systems}\label{Section 3}
In this section, we first introduce the optimum MLD of GSM-AFDM systems. Typically, the complexity of the MLD is excessive, even for a moderate constellation size of $Q$. Therefore, considering large-scale GSM-AFDM systems, we propose four different LMMSE equalizer-based low-complexity detectors. Specifically, we first detail the LMMSE-MLD dividing the equalized symbol vector into $N$ groups and carrying out MLD within each group separately. Then, we design the LMMSE-TC-LLRD to further mitigate the detection complexity. Moreover, based on the greedy compressed sensing algorithms and the GSM codebooks, we propose the GRCD and the RSCD. Finally, the complexity of the above detectors is analyzed. Throughout this section, we neglect the group index $n$ when we decode the $n$th sub-vector for \textcolor{black}{notational} convenience, since the received symbols of each group are processed similarly.
\subsection{Maximum Likelihood Detector}\label{Section3-1}
Upon harnessing the maximum \emph{a posteriori} (MAP) principle and detecting $N$ groups jointly, the optimal symbol detector can be formulated as
\begin{align}
	\pmb{x}^\text{MAPD}=\textcolor{black}{\arg\max_{\pmb{x}\in\Omega}}p(\pmb{x}|\pmb{y}),
\end{align}
where $\Omega=\mathcal{D}^N$ denotes the set of all $2^{NL_b}=2^L$ candidates of the transmit symbol vector $\pmb{x}$. Assuming that all the candidates are independent and equiprobable, then the MAPD can be equivalently expressed as the MLD, formulated as
\begin{align}\label{eq_MLD}
	\pmb{x}^\text{MLD}=\textcolor{black}{\arg\min_{\pmb{x}\in\Omega}}\left\{||\pmb{y}-\pmb{G}\pmb{x}||^2\right\}.
\end{align}
\subsection{LMMSE-MLD and LMMSE-TC-LLRD}\label{Section3-2}
In LMMSE-MLD, we consider performing detection within the $N$ groups separately. The soft estimate $\tilde{\pmb{x}}$ can be obtained based on the LMMSE equalizer relying on \eqref{eq_y}, yielding
\begin{align}\label{eq_LMMSE}
	\tilde{\pmb{x}}=\left(\pmb{G}^H\pmb{G}+\frac{1}{\gamma_s}\pmb{I}_{NM_t}\right)^{-1}\pmb{G}^H\pmb{y}.
\end{align}
By equally partitioning the soft estimate into $N$ sub-vectors, we have $\tilde{\pmb{x}}=[\tilde{\pmb{x}}^T_0,\ldots,\tilde{\pmb{x}}^T_{N-1}]^T$. Then, we focus on the $n$th group. The MLD can be formulated as
\begin{align}\label{eq_MLD_group}
	\pmb{x}^{\text{LMMSE-MLD}}=\textcolor{black}{\arg\min_{\pmb{d}_i\in\mathcal{D}}}\left\|\tilde{\pmb{x}}-\pmb{d}_i\right\|^2,
\end{align}
where all the $2^{L_b}$ candidates in $\mathcal{D}$ are checked. To further reduce the complexity of LMMSE-MLD, we propose the LMMSE-TC-LLRD in the spirit of \cite{bacsar2013orthogonal}. Specifically, upon exploiting the property that the elements of the codebook in \eqref{eq_codebook} can be zero or non-zero, the LLRD can be formulated as
\begin{align}\label{eq_LLRD}
	\lambda(m_t)=\ln\frac{\sum_{q=1}^Q \text{Pr}[x(m_t)=a_q|\tilde{x}(m_t)]}{\text{Pr}[x(m_t)=0|\tilde{x}(m_t)]},
\end{align}
for $m_t=0,\ldots,M_t-1$. A higher value of $\lambda(m_t)$ implies that the TA index $m_t$ is more likely to be active. Based on the GSM property of $\sum_{q=1}^Q \text{Pr}[x(m_t)=a_q]=K/M_t$ and $\text{Pr}[x(m_t)=0]=(M_t-K)/M_t$, \eqref{eq_LLRD} can be rewritten as
\begin{align}\label{eq_LLRD_re}
	\lambda(m_t)&=\ln(K)-\ln(M_t-K)+\frac{|\tilde{x}(m_t)|^2}{N_0}\nonumber\\
	&+\ln\left[\sum_{q=1}^Q\exp\left(-\frac{|\tilde{x}(m_t)-a_q|^2}{N_0}\right)\right].
\end{align}
We sort the LLR values in descending order, yielding
\begin{align}\label{eq_sorting_LLR}
	(i_1,\ldots,i_{M_t}),\ \text{subject to }\lambda(i_1)\geq\ldots\geq\lambda(i_{M_t}).\end{align}
The TA index set $\{i_1,i_2,\ldots,i_K\}$ associated with the $K$ highest LLR values is identified as the preliminary estimate of TAP, which can be further expressed as $\mathcal{G}=\{{\mathcal{G}}_{0},\ldots,{\mathcal{G}}_{K-1}\}$. Since we have $\binom{M_t}{K}>C$, i.e., there are unused TAPs in GSM, the following steps are harnessed for avoiding catastrophic TAP decisions.

The minimum Hamming distance between the specific TAP set $\mathcal{X}$ harnessed and the preliminary estimate TAP $\mathcal{G}$ is denoted as $d_\text{LLR}$. We have $d_\text{LLR}=0$ if $\mathcal{G}\subset\mathcal{X}$, and we have $\mathcal{I}^\text{LMMSE-TC-LLRD}=\mathcal{G}$. Under the scenario of $d_\text{LLR}\neq0$, the $U$ TAPs corresponding to this minimum Hamming distance are collected as TAP candidates, yielding $\tilde{\mathcal{I}}=\left\{\tilde{\mathcal{I}}_1,\ldots,\tilde{\mathcal{I}}_U\right\}\subset\mathcal{X}$. If $U=1$, the first TAP is attained as the final estimated TAP, yielding $\mathcal{I}^\text{LMMSE-TC-LLRD}=\tilde{\mathcal{I}}_1$. When $U>1$, we define a $N_t$-length indicator vector $\pmb{z}_u$ for $u=1,\ldots,U$, whose elements can be formulated as
\begin{align}\label{eq_z}
	z_u(m_t)=\begin{cases}
		1, & \mbox{if }m_t\in\tilde{\mathcal{I}}_u\\
		0, & \mbox{\textcolor{black}{otherwise} }	\end{cases}
\end{align}
Consequently, upon leveraging the TAP candidate set $\tilde{\mathcal{I}}$ and the soft estimate $\tilde{\pmb{x}}$ as \emph{a priori} information, the final estimated TAP ${\mathcal{I}}^\text{LMMSE-TC-LLRD}=\left\{\hat{i}_0,\ldots,\hat{i}_K\right\}$ can be attained as
\begin{align}\label{eq_TAP}
	{\mathcal{I}}^\text{LMMSE-TC-LLRD}=\textcolor{black}{\arg\max_{\tilde{\mathcal{I}}_u\in\tilde{\mathcal{I}}}}||\pmb{z}_u\odot{\pmb{\lambda}}||^2.
\end{align}
Finally, the estimated APM symbol can be obtained by leveraging the symbol-wise ML detection, yielding
\begin{align}\label{eq_hard_ML}
	x_{d}^\text{LMMSE-TC-LLRD}(k)=\textcolor{black}{\arg\underset{a_q\in\mathcal{Q}}\min}\left|\tilde{x}\left(\hat{i}_k\right)-a_q\right|^2.
\end{align}
Our enhanced LMMSE-TC-LLRD is summarized in \textbf{Algorithm \ref{alg-llr}}.
%%%%%%%%%%%%%%%%%%%%%%%%%%%%%%%%%%%%%%%%%%%%%%%%%%%%%%%%%
\begin{algorithm}[htbp]
\footnotesize
%\setstretch{1.5} 
\caption{LMMSE-TC-LLRD}
\label{alg-llr}
\begin{algorithmic}[1]
    \Require
      ${\pmb{y}}$, \textcolor{black}{$\pmb{G}$}, $\mathcal{X}$ and $\gamma_s$.
    \State LMMSE equalization: $\tilde{\pmb{x}}=\left(\pmb{G}^H\pmb{G}+\frac{1}{\gamma_s}\pmb{I}_{NM_t}\right)^{-1}\pmb{G}^H\pmb{y}$.
    \State Equally divide $\tilde{\pmb{x}}$ into $N$ groups as $\tilde{\pmb{x}}=[\tilde{\pmb{x}}^T_0,\ldots,\tilde{\pmb{x}}^T_{N-1}]^T$.
     \Statex $// \textbf{Consider sub-vectors}$ $\pmb{x}_n$ \textbf{for} $n=0,\ldots,N-1$:
    \State $\lambda(m_t)=\ln(K)-\ln(M_t-K)+\frac{|\tilde{x}(m_t)|^2}{N_0}$
    \Statex $+\ln\left[\sum_{q=1}^Q\exp\left(-\frac{|\tilde{x}(m_t)-a_q|^2}{N_0}\right)\right].$
   \State Compute $(i_1,\ldots,i_{M_t}),\ \text{subject to }\lambda(i_1)\geq\ldots\geq\lambda(i_{M_t}).$
   \State Obtain the preliminary estimated TAP as $\mathcal{G}_n=\{{\mathcal{G}}_{0},\ldots,{\mathcal{G}}_{K-1}\}$.
   \State Calculate the minimum Hamming distance $d_\text{LLR}$.
   \If{$d_\text{LLR}=0$}
   \State $\hat{\mathcal{I}}=\mathcal{G}$
   \Else
   \State Identify $U$ TAPs as $\tilde{\mathcal{I}}=\left\{\tilde{\mathcal{I}}_1,\ldots,\tilde{\mathcal{I}}_U\right\}\subset\mathcal{X}$.
   \If{$U=1$}
   \State $\hat{\mathcal{I}}=\tilde{\mathcal{I}}_1$.
   \Else
   \State $z_u(m_t)=\begin{cases}
		1, & \mbox{if }m_t\in\tilde{\mathcal{I}}_u\\
		0, & \mbox{if }m_t\notin\tilde{\mathcal{I}}_u.	\end{cases}$
\State $\hat{\mathcal{I}}=\left\{\hat{i}_0,\ldots,\hat{i}_K\right\}=\textcolor{black}{\arg\max\limits_{\tilde{\mathcal{I}}_u\in\tilde{\mathcal{I}}}}||\pmb{z}_u\odot{\pmb{\lambda}}||^2.$
   \EndIf
   \EndIf
\State $x_{d}^\text{LMMSE-TC-LLRD}(k)=\textcolor{black}{\arg\underset{a_q\in\mathcal{Q}}\min}\left|\tilde{x}\left(\hat{i}_k\right)-a_q\right|^2$.
\State \textbf{Output} $\mathcal{I}^{\text{LMMSE-TC-LLRD}}=\hat{\mathcal{I}}$ and ${\pmb{x}}_{d}^{\text{LMMSE-TC-LLRD}}$.
\end{algorithmic}
\end{algorithm}
\subsection{Greedy Residual Check Detector (GRCD)}\label{Section3-3}
In GRCD, we intend to reduce the complexity by detecting the TAP-indices and classical APM symbols separately. The philosophy of greedy algorithms is invoked to find the local optimum during each iteration. Specifically, the GRCD first estimates the reliability of each element in the soft estimate $\tilde{\pmb{x}}$ and then performs iterative detection by checking the residual signal corresponding to different TAPs. The details of our GRCD are illustrated below.

During the reliability estimation stage, we also invoke the LMMSE equalizer of \eqref{eq_LMMSE} to attain the $n$th soft estimate $\tilde{\pmb{x}}_n$. Typically, it is observed that the elements of $\tilde{\pmb{x}}_n$ having high magnitudes correspond to the active elements with high probability, yielding high reliability, particularly in high-SNR scenarios. Consequently, similar to \eqref{eq_sorting_LLR}, we carry out reliability estimation by sorting the element indices in descending order based on their magnitudes. The sorted indices can be formulated as $(l_{1},\ldots,l_{M_t})$, where $l_{q}\in\{0,\ldots,M_t-1\}$ for $q=1,\ldots,M_t$, and we have $l_{q}\neq l_{j}, \forall q\neq j$.

Next, our GRCD enters the iterative detection stage. During the $t$th iteration, the TA index $l_{t}$ is tested with the highest priority. Specifically, $C_{t}$ TAPs that include the TA index $l_{t}$ are selected from the set $\mathcal{X}$, which can be expressed as $\mathcal{X}_{t}=\{\mathcal{X}_{t}^1,\ldots,\mathcal{X}_{t}^{C_{t}}\}\subset\mathcal{X}$, where we have $\bigcap_{c_{t}=1}^{C_{t}}\mathcal{X}^{c_{t}}_{t}=l_{t}$. Then, the APM symbol detection is processed by exploiting $\mathcal{X}_{t}$ as the $\emph{a priori}$ information, yielding
\begin{align}
	\tilde{\pmb{x}}_{d,{c_t}}=\textcolor{black}{\arg\min_{\pmb{a}\in\mathbb{C}^{K}}}\left\|\tilde{\pmb{x}}-\pmb{\Upsilon}_{\mathcal{X}_{t}^{c_t}}\pmb{a}\right\|^2.
\end{align}
%%%%%%%%%%%%%%%%%%%%%%%%%%%%%%%%%%%%%%%%%%%%%%%%%%%%%%%%%
\begin{algorithm}[t]
\footnotesize
%\setstretch{1.5} 
\caption{Greedy Residual Check Detector}
\label{alg-grcd}
\begin{algorithmic}[1]
    \Require
      ${\pmb{y}}$, \textcolor{black}{$\pmb{G}$}, $\mathcal{X}$, $\gamma_s$ and $\epsilon_{\text{th}}$.
      \State \textbf{Preparation}: Set the maximum number of iteration $T_1$, $\mathcal{X}_\text{check}=\mathcal{X}$, $\epsilon_{\infty}=\infty$, $\mathcal{I}^{\text{GRCD}}=\emptyset$ and $\pmb{x}_d^{\text{GRCD}}=\emptyset$.
    \Statex $// \textbf{Reliability Estimation:}$    
    \State \textcolor{black}{$\tilde{\pmb{x}}=\left(\pmb{G}^H\pmb{G}+\frac{1}{\gamma_s}\pmb{I}_{NM_t}\right)^{-1}\pmb{G}^H\pmb{y}$.}
    \Statex $// \textbf{Consider sub-vectors}$ $\pmb{x}_n$ \textbf{for} $n=0,\ldots,N-1$:   
    \State $\mathcal{L}=(l_1,\ldots,l_{M_t}),\text{subject to }\left|\tilde{x}(l_1)\right|^2\geqslant\ldots\geqslant\left|\tilde{x}(l_{M_t})\right|^2$.
    \Statex $// \textbf{Iterative Detection:}$  
    \For{$t=1$ to $T_1$}
    \State $\mathcal{X}_t=\{\mathcal{X}_t^1,\ldots,\mathcal{X}_t^{C_t}\}\subset\mathcal{X}_\text{check}$, where $\bigcap_{c_t=1}^{C_t}\mathcal{X}^{c_t}_t=l_t$.
    \If{$\mathcal{X}_t=\emptyset$}
    \BREAK
    \Else
    \For{$c_t=1$ to $C_t$}
    \State $\tilde{\pmb{x}}_{d,c_t}=\pmb{\Upsilon}_{\mathcal{X}^{c_t}_t}^{\dagger}{\tilde{\pmb{x}}}$,
    \State $a_{d,c_t}(k)=\textcolor{black}{\arg\underset{a_q\in\mathcal{Q}}\min}\left|\tilde{{x}}_{d,c_t}(k)-a_q\right|^2$, 
    \Statex\quad\quad\quad\quad\quad \textcolor{black}{$\forall k=0,\ldots,K-1$.}
    \EndFor
    \State 
    	$\left(\mathcal{I}_t,{\pmb{x}}_{d,t}\right)=\textcolor{black}{\arg\min\limits_{\mathcal{X}_t^{c_t}\subset\mathcal{X}^t,\pmb{a}_{d,c_t}\in\mathcal{A}_t}}\left\|\tilde{\pmb{x}}-\pmb{\Upsilon}_{\mathcal{X}_t^{c_t}}\pmb{a}_{d,c_t}\right\|^2$.
	\State $\epsilon_t=\left\|\tilde{\pmb{x}}-\pmb{\Upsilon}_{\mathcal{I}_t}\pmb{x}_{d,t}\right\|^2$.
	\If{$\epsilon_t<\epsilon_{\text{th}}$}
	\State $\mathcal{I}^{\text{GRCD}}=\mathcal{I}_t$ and $\pmb{x}_d^{\text{GRCD}}=\pmb{x}_{d,t}$,
	\BREAK
	\ElsIf{$\epsilon_t<\epsilon_{\infty}$}
	\State $\mathcal{X}_\text{check}\leftarrow\mathcal{X}_\text{check}\setminus\mathcal{X}_t$, $\epsilon_{\infty}=\epsilon_{t}$, $\mathcal{I}^{\text{GRCD}}=\mathcal{I}_t$ and
	\Statex \qquad\quad\quad\quad $\pmb{x}_d^{\text{GRCD}}=\pmb{x}_{d,t}$.
	\EndIf
	\EndIf
    \EndFor
\State \textbf{Output} $\mathcal{I}^{\text{GRCD}}=\mathcal{I}_t$ and $\pmb{x}_d^{\text{GRCD}}=\pmb{x}_{d,t}$.
\end{algorithmic}
\end{algorithm}
Upon utilizing the popular least square technique, the solution $\tilde{\pmb{x}}_{d,c_t}=\pmb{\Upsilon}_{\mathcal{X}^{c_t}_t}^{\dagger}{\tilde{\pmb{x}}}$ can be formulated as
\begin{align}\label{eq_LS}
	\tilde{\pmb{x}}_{d,c_t}=\pmb{x}_d+\pmb{r}_{\mathcal{X}^{c_t}_t,\mathcal{I}}+\tilde{\pmb{n}},
\end{align}
where the residual detection error under the scenario that the TAP is detected correctly, i.e., we have $\pmb{r}_{\mathcal{X}^{c_t}_t,\mathcal{I}}=\pmb{0}$ when $\mathcal{X}^{c_t}_t=\mathcal{I}$, and $\tilde{\pmb{n}}=\pmb{\Upsilon}_{\mathcal{X}^{c_t}_t}^{\dagger}{\pmb{n}}$ represents the AWGN vector. Then, the estimates of the APM symbols ${\pmb{a}}_{d,c_t}=[a_{d,c_t}(0),\ldots,a_{d,c_t}(K-1)]^T$ can be obtained by considering the symbol-wise ML detection of \eqref{eq_hard_ML}. Denoting all the $C_t$ estimated APM symbol sets as $\mathcal{A}_t=\{{\pmb{a}}_{d,1},\ldots,{\pmb{a}}_{d,C_t}\}$, the optimum can be obtained as
\begin{align}\label{eq_opt}
	\left(\mathcal{I}_t,{\pmb{x}}_{d,t}\right)=\textcolor{black}{\arg\min\limits_{\mathcal{X}_t^{c_t}\subset\mathcal{X}^t,\pmb{a}_{d,c_t}\in\mathcal{A}_t}}\left\|\tilde{\pmb{x}}-\pmb{\Upsilon}_{\mathcal{X}_t^{c_t}}\pmb{a}_{d,c_t}\right\|^2.
\end{align}
Therefore, the residual error is given by $\epsilon_t=\left\|\tilde{\pmb{x}}-\pmb{\Upsilon}_{\mathcal{I}_t}\pmb{x}_{d,t}\right\|^2$. Finally, the proposed GRCD updates the detection results when $\epsilon_t<\epsilon_{t-1}$, and the iteration terminates if $\epsilon_t\leq\epsilon_{\text{th}}$ or the maximum number of iterations is reached. The GRCD is summarized in \textbf{Algorithm  \ref{alg-grcd}}. \textcolor{black}{We emphasize that the step $\mathcal{X}_\text{check}\leftarrow\mathcal{X}_\text{check}\setminus\mathcal{X}_t$ of Line 19 in \textbf{Algorithm \ref{alg-grcd}} implies that the $(t+1)$th GRCD iteration can avoid testing the TAPs in the set $\mathcal{X}_t$ associated with the $t$th GRCD iteration.}
\subsection{Reduced-space Check Detector (RSCD)}\label{Section3-4}
Next, we carry out the detections of TAP-indices and APM symbol separately in the RSCD. Our RSCD can attain a near-MMSE-ML BER performance by only checking a reduced TAP space. However, the reliabilities of all elements in $\tilde{\pmb{x}}$ are used in contrast to the GRCD. Our RSCD first employs the LMMSE detection of \eqref{eq_LMMSE} to obtain the soft estimate $\tilde{\pmb{x}}$ and the corresponding hard-decision vector $\hat{\pmb{x}}$. Focusing on the $n$th group, our RSCD attains the element reliability metrics according to the TAP space $\mathcal{X}$, which can be formulated as
\begin{align}\label{eq_metric}
	\alpha_c=\sum_{k=0}^{K-1}\left|\hat{x}[\mathcal{X}_c(k)]-\tilde{x}[\mathcal{X}_c(k)]\right|^2,\ c=1,\ldots,C.
\end{align}
 Consequently, we sort the reliabilities of all the TAPs in ascending order, yielding
\begin{align}\label{eq_relia}
	\mathcal{R}=({r_1},\ldots,{r_C}),\quad\text{subject to }{\alpha}_{r_1}\leqslant\ldots\leqslant{\alpha}_{r_C},
\end{align}
where $r_c\in\{1,\ldots,C\},\forall c$ and we have ${\alpha}_{r_i}\neq{\alpha}_{r_j}$, $\forall i\neq j$. Similarly, we consider the TAP associated with a higher value of ${\alpha}_{r_c}$ to be the correct detection that results in a higher probability, which is more evident within the high-SNR region. The RSCD first checks the TAP concerning the reliability metric ${\alpha}_{r_1}$ in the following APM symbol detection stage, and the intricate details are introduced as follows.

During the $t$th iteration of the second stage, the RSCD first selects the TAP $\mathcal{I}_t$ corresponding to the reliability metric $\alpha_{r_t}$. Then, the classic APM symbols $\pmb{x}_{d,t}\in\mathcal{A}^{K}$ can be detected upon leveraging the LS approach and the symbol-wise ML detection characterized in \eqref{eq_hard_ML} and \eqref{eq_LS}, respectively. Then, our RSCD groups the detected TAP set and APM symbols as ${\mathcal{D}}_t=\{\mathcal{I}_t,\pmb{x}_{d,t}\}$. The detection of residual error can be attained similarly to the process of our GRCD. The number of checked TAPs is denoted as $T_2$, i.e., the proposed RSCD employs $T_2$ iterations during the second stage. Assuming that the $t_\text{opt}$th iteration attains the minimum residual, we have
\begin{align}
	\textcolor{black}{t_\text{opt}=\arg\min\limits_{t}\{\epsilon_t|\forall t=1,\ldots,{T_2}\}.}
\end{align}
Finally, the optimal detection can be formulated as $(\mathcal{I}^{\text{RSCD}},\pmb{x}_d^{\text{RSCD}})=(\mathcal{I}_{t_\text{opt}},\pmb{x}_{d,t_\text{opt}})$. The proposed RSCD is summarized in \textbf{Algorithm \ref{alg-rscd}}.
%%%%%%%%%%%%%%%%%%%%%%%%%%%%%%%%%%%%%%%%%%%%%%%%%%%%%%%%%
\begin{algorithm}[htbp]
\footnotesize
%\setstretch{1.5} 
\caption{Reduced Space Check Detector}
\label{alg-rscd}
\begin{algorithmic}[1]
    \Require
      ${\pmb{y}}$, \textcolor{black}{$\pmb{G}$}, $\mathcal{X}$ and $\gamma_s$.
      \State \textbf{Preparation}: Set the maximum number of iteration $T_2$.
    \Statex $// \textbf{Reliability Estimation:}$    
    \State Carry out LMMSE detection as
    \State $\tilde{\pmb{x}}=\left(\pmb{G}^H\pmb{G}+\frac{1}{\gamma_s}\pmb{I}_{NM_t}\right)^{-1}\pmb{G}^H\pmb{y}$.
    \Statex $// \textbf{Consider the}\ n\text{th group:}$ 
    \State Calculate the TAP reliability metrics as
    \State \textcolor{black}{$\alpha_c=\sum_{k=0}^{K-1}\left|\hat{x}[\mathcal{X}_c(k)]-\tilde{x}[\mathcal{X}_c(k)]\right|^2,\ c=1,\ldots,C.$}
    \State Sort the TAP reliabilities as
    \State \textcolor{black}{$\mathcal{R}=({r_1},\ldots,{r_C}),\quad\text{subject to }{\alpha}_{r_1}\leqslant\ldots\leqslant{\alpha}_{r_C}$.}
    \Statex $// \textbf{Iterative Detection:}$  
    \For{$t=1$ to $T_2$}
    \State Collect the TAP $\mathcal{L}_t$ according to $\alpha_{r_t}$.
    \State \textcolor{black}{Obtain estimated APM symbols as $\tilde{\pmb{x}}_{d,t}=\pmb{\Upsilon}_{\mathcal{L}_t}^{\dagger}\tilde{\pmb{x}}$,}
    \State $x_{d,c_t}(k)=\textcolor{black}{\arg\underset{a_q\in\mathcal{Q}}\min}\left|\tilde{{x}}_{d,c_t}(k)-a_q\right|^2$, 
    \State for $k=0,\ldots,K-1$.    
    \State Collect detected results as $\{\mathcal{I}_t,\pmb{x}_{d,t}\}$.
    \State Attain the residual error as $\epsilon_t=\left\|\tilde{\pmb{x}}-\pmb{\Upsilon}_{\mathcal{I}_t}\pmb{x}_{d,t}\right\|^2$.
    \EndFor
    \State Obtain the final index as \textcolor{black}{$t_\text{opt}=\arg\min\limits_{t}\{\epsilon_t|\forall t=1,\ldots,{T_2}\}$}.
\end{algorithmic}
\end{algorithm}
\subsection{Complexity Analysis}\label{Section3-5}
According to \eqref{eq_MLD}, all the $2^L$ possible candidates in $\Omega$ are tested. Therefore, the complexity of the MLD is on the order of $\mathcal{O}(2^L)$. 

\textcolor{black}{Our analysis in Subsection \ref{Section3-2} shows that the complexity of our LMMSE-MLD is given by that of the LMMSE equalizer and MLD. Explicitly, based on popular matrix decomposition methods, the complexity of the LMMSE equalizer can be expressed as $\mathcal{O}(N^2 M_t^2)$ \cite{tuchler2002minimum}. The MLD is performed within $N$ groups, whereby the complexity of each group is $\mathcal{O}(2^{L_b})$. Therefore, the overall complexity of LMMSE-MLD can be expressed as $\mathcal{O}(N^2 M_t^2+N2^{L_b})$.}

\textcolor{black}{The complexity of calculating \eqref{eq_LLRD_re} is $\mathcal{O}(Q)$. Moreover, the complexity of calculating  \eqref{eq_TAP} and \eqref{eq_hard_ML} can be respectively expressed by $\mathcal{O}(M_tU)$ and $\mathcal{O}(QK)$. Therefore, the total complexity of our LMMSE-TC-LLRD can be expressed as $\mathcal{O}[N^2 M_t^2+N(M_tQ+M_tU+QK)]$.}

\textcolor{black}{In \textbf{Algorithm \ref{alg-grcd}}, the complexities of Lines 2, 3, 11, and 13 are given as $\mathcal{O}(N^2M_t^2)$, $\mathcal{O}(M_t\log M_t)$, $\mathcal{O}(QK)$ and $\mathcal{O}(M_t)$, respectively. Moreover, we can observe from Subsection \ref{Section3-3} that the complexity of the proposed GRCD is dominated by the number of TAP candidates of each iteration and the number of iterations. Specifically, the best case is when $T_1=1$, i.e., the proposed GRCD terminates after a single iteration and only a single TAP is checked. Under this scenario, the complexity can be expressed as $\mathcal{O}[N^2 M_t^2+N(M_t\log M_t+QK+M_t)]$. By contrast, we have the worst case of checking all the $2^{L_1}$ TAPs. In this case, it can be readily shown that the corresponding complexity is $\mathcal{O}[N^2 M_t^2+N(M_t\log M_t+2^{L_1}(QK+M_t))]$. We emphasize that the complexity of the worst case is still lower than that of the MLD, since our GRCD employs the simple symbol-wise ML detection of \eqref{eq_hard_ML}. Typically, as shown in our simulation results of Section \ref{Section 5}, only $T_1<C=2^{L_1}$ TAPs have to be checked. Therefore, the total detection complexity is given by $\mathcal{O}[N^2 M_t^2+N(M_t\log M_t+T_1(QK+M_t))]$.}

\textcolor{black}{Based on Subsection \ref{Section3-4}, the complexity of each RSCD iteration is $\mathcal{O}(QK+M_t)$. Moreover, the complexities of Lines 5 and 7 of \textbf{Algorithm \ref{alg-rscd}} can be formulated as $\mathcal{O}(CK)$ and $\mathcal{O}(C\log C)$, respectively. Therefore, by considering the LMMSE equalizer, the overall RSCD complexity is given by $\mathcal{O}[N^2 M_t^2+N(CK+C\log C+T_2(QK+M_t)]$. Consequently, the worst case is attained when $T_{2,\text{max}}=2^{L_1}$ iterations are employed, yielding a complexity of $\mathcal{O}[N^2 M_t^2+N(CK+C\log C+2^{L_1}(QK+M_t)]$. However, since the reliabilities of all the elements in $\hat{\pmb{x}}$ are leveraged, the proposed RSCD can achieve near-LMMSE-MLD performance by only testing a fraction space of the TAP space $\mathcal{X}$, as introduced in Section \ref{Section 5}.}
\vspace{-1em}
\section{Performance Analysis}\label{Section 4}
\subsection{Analysis of BER Performance}\label{Section4-1}
The input-output relationship of \eqref{eq_SISO} can be rewritten as
\begin{align}\label{eq_y_SISO_re}
	\pmb{y}_{m_r,m_t}&=\sum_{p=1}^P h_{p,m_r,m_t} \pmb{H}_p\pmb{z}_{m_t}+\textcolor{black}{\pmb{w}_{m_r,m_t}}\nonumber\\
	&=\pmb{\Xi}(\pmb{z}_{m_t})\pmb{h}_{m_r,m_t}+\textcolor{black}{\pmb{w}_{m_r,m_t}},
\end{align}
where the $N\times P$-dimensional concatenated matrix is given by $\pmb{\Xi}(\pmb{z}_{m_t})=[\pmb{H}_1\pmb{z}_{m_t}|\ldots|\pmb{H}_P\pmb{z}_{m_t}]$, and $\pmb{h}_{m_r,m_t}=[h_{1,m_r,m_t},\ldots,h_{P,m_r,m_t}]^T\in\mathbb{C}^P$ is the channel coefficient vector. Let us denote the stacked equivalent channel matrix and the channel coefficient vector respectively by ${\pmb{\Xi}}(\pmb{z})=[\pmb{\Xi}(\pmb{z}_{0}),\ldots,\pmb{\Xi}(\pmb{z}_{M_t -1})]\in\mathbb{C}^{N\times PM_t}$ and $\pmb{h}_{m_r}=[\pmb{h}_{m_r,0}^T,\ldots,\pmb{h}_{m_r,M_t-1}^T]^T\in\mathbb{C}^{PM_t}$. Then, the $m_r$th received signal $\pmb{y}_{m_r}$ can be rewritten as $\pmb{y}_{m_r}=\pmb{\Xi}(\pmb{z})\pmb{h}_{m_r}+\textcolor{black}{\pmb{w}_{m_r}}$. Upon introducing $\pmb{\Psi}(\pmb{z})=\pmb{I}_{M_r}\otimes\pmb{\Xi}(\pmb{z})\in\mathbb{C}^{NM_r\times PM_t M_r}$ and ${\pmb{h}}=[\pmb{h}_0^T,\ldots,\pmb{h}_{M_r-1}^T]^T\in\mathbb{C}^{PM_t M_r}$, the DAFT-domain end-to-end input-output relationship of \eqref{eq_y} can be expressed as
\begin{align}\label{eq_y_per}
	\pmb{y}=\pmb{\Psi}(\pmb{z}){\pmb{h}}+\textcolor{black}{\pmb{w}}.
\end{align}
Consequently, the MLD associated with the equivalent end-to-end input-output relationship of \eqref{eq_y_per} can be reformulated as $\pmb{z}^\text{ML}=\argmin\limits_{\pmb{f}_i\in\Omega}\left\{\left\|\pmb{y}-\pmb{\Psi}(\pmb{f}_i){\pmb{h}}\right\|^2\right\}$. Upon considering the pairwise error event $\{\pmb{z}^c,\pmb{z}^e\}$, where $\pmb{z}^c=\pmb{f}_i$ represents the transmit codeword vector and $\pmb{z}^e=\pmb{f}_j$ denotes the erroneous detected codeword vector associated with $\forall i\neq j$, i.e., we have $\pmb{f}_i\neq\pmb{f}_j$ for $\pmb{f}_i,\pmb{f}_j\in\mathcal{S}$. Furthermore, let us define the error vector space as $\mathcal{E}=\{\pmb{e}=\pmb{f}_i-\pmb{f}_j,\forall \pmb{f}_i,\pmb{f}_j\in\Omega,\forall i\neq j\}$. Bearing in mind that $\pmb{y}=\pmb{\Psi}(\pmb{z}^c){\pmb{h}}+\textcolor{black}{\pmb{w}}$ with a given ${\pmb{h}}$, the conditional \textcolor{black}{pairwise error probability (PEP)} can be formulated based on $\pmb{\Psi}(\pmb{e})=\pmb{\Psi}(\pmb{z}^e)-\pmb{\Psi}(\pmb{z}^c)$ as $P(\pmb{z}^c,\pmb{z}^e|\pmb{h})=\text{Pr}\left[\Re\left\{\pmb{w}^H\pmb{\Psi}(\pmb{e}){\pmb{h}}\right\}\geq{\left\|\pmb{\Psi}(\pmb{e}){\pmb{h}}\right\|^2}/{2}\right]$. \textcolor{black}{Given $\pmb{e}$ and $\pmb{h}$, it can be readily shown that $\pmb{w}^H\pmb{\Psi}(\pmb{e}){\pmb{h}}$ has the mean of zero, considering the fact that the AWGN vector obeys $\pmb{w}\sim\mathcal{CN}(\pmb{0},\frac{1}{\gamma_s}\pmb{I}_{NM_r})$. The corresponding variance can be formulated as
\begin{align}
	\sigma_w^2&=\mathbb{E}\{\pmb{w}^H\pmb{\Psi}(\pmb{e})\pmb{h}\pmb{h}^H\pmb{\Psi}(\pmb{e})^H\pmb{w}\}\nonumber\\
	&=\mathbb{E}\{\tr[\pmb{\Psi}(\pmb{e})\pmb{h}\pmb{h}^H\pmb{\Psi}(\pmb{e})^H\pmb{w}\pmb{w}^H]\}\nonumber\\
	&=\tr[\pmb{\Psi}(\pmb{e})\pmb{h}\pmb{h}^H\pmb{\Psi}(\pmb{e})^H\mathbb{E}\{\pmb{w}\pmb{w}^H\}]\nonumber\\
	&=\frac{\left\|\pmb{\Psi}(\pmb{e}){\pmb{h}}\right\|^2}{\gamma_s}.	\end{align}
Therefore, we have $\Re\left\{\pmb{w}^H\pmb{\Psi}(\pmb{e}){\pmb{h}}\right\}\sim\mathcal{N}\left(0,\frac{\left\|\pmb{\Psi}(\pmb{e}){\pmb{h}}\right\|^2}{2\gamma_s}\right)$.} Upon letting \textcolor{black}{$\chi=\left\|\pmb{\Psi}(\pmb{e}){\pmb{h}}\right\|^2$}, we have
\begin{align}\label{eq_PEP3}	
	P(\pmb{z}^c,\pmb{z}^e|\pmb{h})=Q\left(\sqrt{\frac{\gamma_s}{2}\textcolor{black}{\chi}}\right),
\end{align}
where $Q(x)$ denotes the Gaussian $Q$-function. For $x>0$, we have $Q(x)=\frac{1}{\pi}\int_{0}^{\pi/2} \exp\left(-\frac{x^2}{2\sin^2\theta}\right)d\theta$ \cite{6045355}. Alternatively, \eqref{eq_PEP3} can be expressed as
\begin{align}\label{eq_PEP4}		
P(\pmb{z}^c,\pmb{z}^e|\pmb{h})=\frac{1}{\pi}\int_{0}^{\frac{\pi}{2}}\exp\left(-\frac{\gamma_s\textcolor{black}{\chi}}{4\sin^2\theta}\right)d\theta.
\end{align}
\textcolor{black}{The unconditional pairwise error probability (UPEP) can be attained upon averaging $P(\pmb{z}^c,\pmb{z}^e|\pmb{h})$ with respect to the distribution of $\chi$, yielding \cite{6387987}
\begin{align}\label{eq_UPEP1}
	P(\pmb{z}^c,\pmb{z}^e)&=\frac{1}{\pi}\int_0^{\frac{\pi}{2}}\int_0^{\infty}\exp\left(-\frac{\gamma_s\chi}{4\sin^2\theta}\right)p_\chi(\chi)d\chi d\theta\nonumber\\
	&=\frac{1}{\pi}\int_{0}^{\frac{\pi}{2}}{\Phi}_{\chi}\left(-\frac{\gamma_s}{4\sin^2 \theta}\right)d\theta,
\end{align}
where ${\Phi}_{\chi}(t)\triangleq\int_0^\infty \exp(\chi t)p_\chi(\chi)d\chi$ denotes the moment generating function (MGF) with respect to $\chi$, and $p_\chi(\chi)$ denotes the PDF of $\chi$.} It can be shown that $\chi=\left\|\pmb{\Psi}(\pmb{e}){\pmb{h}}\right\|^2=\pmb{h}^H\left[\pmb{\Psi}(\pmb{e})\right]^H\pmb{\Psi}(\pmb{e})\pmb{h}=\pmb{h}^H(\pmb{I}_{M_r}\otimes\pmb{R})\pmb{h}$ along with $\pmb{R}=\left[\pmb{\Xi}(\pmb{e})\right]^H\pmb{\Xi}(\pmb{e})$. Let us assume that elements in $\pmb{h}$ obey a Gaussian distribution with zero mean and variance of $1/(2P)$ per real dimension. Based on the technique in \cite{979326}, the MGF can be expressed as ${\Phi}_{\chi}(t)=\det\left[\pmb{I}_{P M_t M_r}-{t(\pmb{I}_{M_r}\otimes\pmb{R})}/{P}\right]^{-1}$. By defining the rank and non-zero eigenvalues of $\pmb{R}$, respectively, as $r=\text{rank}(\pmb{R})$ and $\{\lambda_1,\ldots,\lambda_r\}$, the UPEP can be formulated as
\begin{align}\label{eq_UPEP3}
	P(\pmb{z}^c,\pmb{z}^e)=\frac{1}{\pi}\int_0^{\frac{\pi}{2}}\left[\prod_{i=1}^{r}\left(1+\frac{\lambda_i\gamma_s}{4P\sin^2\theta}\right)\right]^{-M_r}d\theta.
\end{align}
Since we have ${\lambda_j\gamma}/{(4P\sin^2\theta)}\geqslant{\lambda_j\gamma}/{4P}$ in \eqref{eq_UPEP3}, by considering high-SNR cases having $\gamma_s\gg 1$, we arrive at
\begin{align}\label{eq_UPEP2-2}
	P(\pmb{z}^c,\pmb{z}^e)\leqslant\frac{1}{2}\left[\left(\prod_{i=1}^{r}\lambda_i\right)^{1/r}\left(\frac{\gamma_s}{4P}\right)\right]^{-r M_r}.	
	\end{align}
Finally, upon harnessing the union-bound technique, the average BER of the GSM-AFDM system can be approximated as
\begin{align}\label{eq_ABER}
	P_\text{avg}\approx\frac{1}{2^L L}\sum_{\pmb{b}^c\neq\pmb{b}^e}\xi(\pmb{b}^c,\pmb{b}^e)P(\pmb{z}^c,\pmb{z}^e),
\end{align}
where $\pmb{b}^c$ and $\pmb{b}^e$ denote the corresponding bit sequences of $\pmb{z}^c$ and $\pmb{z}^e$, while $\xi(\cdot,\cdot)$ is the Hamming distance operator between two bit sequences. It should be noted that the BER upper-bound of \eqref{eq_ABER} is also valid for MIMO-AFDM,  since MIMO-AFDM can be considered as a special case of GSM-AFDM.
\subsection{Diversity Order, Coding Gain and DCMC Capacity}\label{Section4-2}
The exponent term of \eqref{eq_UPEP2-2} represents the diversity order upon utilizing the MLD, yielding
\begin{align}\label{eq_diversity}
	V_D=\underset{\forall \pmb{e}\in\mathcal{E}}\min rM_r=\underset{\forall \pmb{e}\in\mathcal{E}}\min\text{rank}(\pmb{R})M_r.
\end{align}
Furthermore, the coding gain can be formulated as $V_C=\underset{\forall \pmb{e}\in\mathcal{E}}\min\left(\prod_{i=1}^{r}\lambda_i\right)^{1/r}.$ It can be observed from \eqref{eq_UPEP2-2} that the diversity order $V_D$ determines the decay rate of our derived UPEP upon increasing the SNR. In addition, the horizontal shift of the UPEP from the baseline $({\gamma_s}/{4P})^{-V_D}/2$ is dominated by the coding gains $V_C$.

Since different TAs transmit independent symbol vectors, it can be readily shown that we have $\text{rank}(\pmb{R})=\text{rank}(\pmb{\Xi}(\pmb{e}))=\text{rank}(\pmb{\Xi}(\pmb{e}_{m_t})),\forall m_t$ with $\pmb{\Xi}(\pmb{e}_{m_t})=\pmb{\Xi}(\pmb{z}^c_{m_t})-\pmb{\Xi}(\pmb{z}^e_{m_t})$. Moreover, it has been derived in \cite{10087310} that we have $\min_{\pmb{z}^c_{m_t}\neq\pmb{z}^c_{m_r}}\text{rank}[\pmb{\Xi}(\pmb{e}_{m_t})]=P$, when the AFDM parameter $c_1$ is set as in \eqref{eq_c1} and the number of subcarriers $N$ satisfies \eqref{eq_N}. Hence, our GSM-AFDM system can attain full diversity associated with $V_D=PM_r$ using MLD and $V_D=P(M_t-M_r+1)$ based on the LMMSE equalizer \cite{4359528}.

Based on \eqref{eq_codebook} and \eqref{eq_y}, the DCMC capacity can be expressed as \cite{1608632}
\begin{align}\label{eq_DCMC1}	
C_{\text{D}}&=\frac{1}{N}\underset{p(\pmb{f}_i)}\max\sum_{i=1}^{2^L}\int_{-\infty}^{\infty}\ldots\int_{-\infty}^{\infty}p({\pmb{y}}|\pmb{f}_i)p(\pmb{f}_i)\bar{p}_i d{\pmb{y}},
	\end{align}
where $\bar{p}_i=\log_2\left[{p({\pmb{y}}|\pmb{f}_i)}/{\sum_{j=1}^{2^L}p({\pmb{y}}|\pmb{f}_j)p(\pmb{f}_j)}\right]$ and $p({\pmb{y}}|\pmb{f}_i)$ is shown in \eqref{eq_pdf}, given that $\pmb{f}_i$ denotes the transmit signal. Since all the candidates in the codebook $\Omega$ are independent and equiprobable associated with $p(\pmb{f}_i)=1/2^L,\forall i$, we have $\bar{p}_i=L-\log_2\sum_{j=1}^{2^L}\exp(\Theta_{i,j})$, where $\Theta_{i,j}=\gamma_s\left[-||\pmb{G}(\pmb{f}_i-\pmb{f}_j)+\textcolor{black}{\pmb{w}}||^2+||\textcolor{black}{\pmb{w}}||^2\right]$. Therefore, the DCMC capacity of the proposed GSM-AFDM scheme can be formulated as
\begin{align}\label{eq_DCMC2}
	C_{\text{D}}&=\frac{L}{N}-\frac{1}{N2^L}\sum_{i=1}^{2^L}\mathbb{E}_{\pmb{G}}\left[\log_2\sum_{j=1}^{2^L}\exp\left(\Theta_{i,j}\right)\right],
\end{align}
where the expectation can be calculated by invoking the Monte Carlo averaging method.
\section{Simulation Results}\label{Section 5}
In this section, we evaluate the performance of the proposed GSM-AFDM systems. The GSM-based and SM-based systems are parameterized by the sets $(M_t,M_r,N,K,Q)$ and $(M_t,M_r,N,Q)$, respectively, while the MIMO-AFDM systems are characterized by $(M_t,M_r,N,Q)$. We first evaluate the BER performance of MLD, the BER upper bound and the DCMC capacity derived. Unless specifically defined, we set the maximum speed as \textcolor{black}{$v=540$ km/h}, while the carrier frequency and the carrier spacing are $f_c=4$ GHz and $\Delta f=2$ kHz, respectively, yielding the normalized maximum Doppler shift of $\alpha_\text{max}=1$. The normalized maximum delay shift is $l_\text{max}=P-1$, while the $p$th normalized delay indices are set as $l_1=0$ and $l_p\in\mathcal{U}[1,l_\text{max}]$, $\forall p\neq 1$. The channel gain coefficients are set as $h_p\sim\mathcal{CN}(0,1/P)$, $\forall p$ \cite{10183832}. Based on Jake's spectrum, the normalized Doppler shifts of the $p$th paths are generated as $k_p=k_{max}\cos(\phi_p)$ with $\phi_p\in\mathcal{U}[-\pi,\pi]$.

\begin{figure}[htbp]
\vspace{-1em}
\centering
\includegraphics[width=0.9\linewidth]{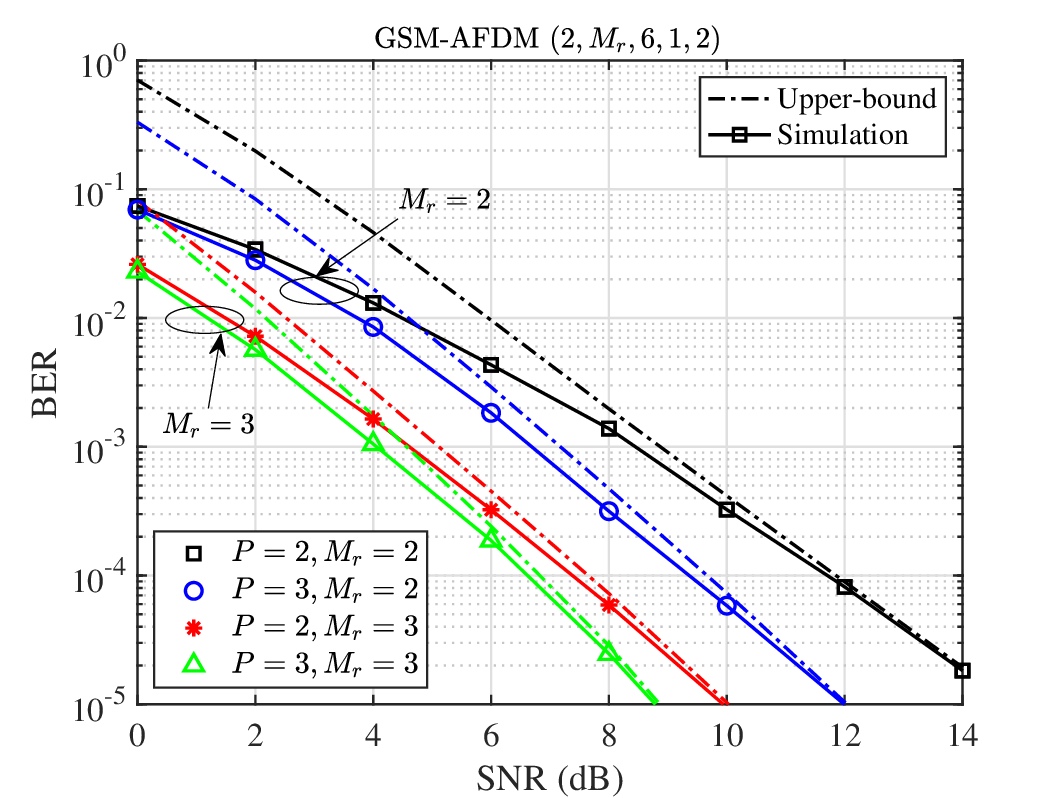}
\caption{BER comparison of MLD and upper-bounds for GSM-AFDM $(2,M_r,6,1,2)$ with \textcolor{black}{$M_r=\{2,3\}$} and $P=\{2,3\}$ at the data rate of $2$ bits/s/subcarrier, where the upper-bounds are computed based on \eqref{eq_ABER}.}
\label{Figure3}
\vspace{-1em}
\end{figure}

In Fig. \ref{Figure3}, we investigate the BER performance of MLD and our BER upper-bounds derived for GSM-AFDM $(2,M_r,6,1,2)$ shown in \eqref{eq_ABER}. Specifically, \textcolor{black}{$M_r=\{2,3\}$} RAs and high-mobility channels having $P=\{2,3\}$ paths and only integer Doppler shifts are considered. Furthermore, $l_\text{max}=1$ is used. From Fig. \ref{Figure3}, we obtain the following observations. Firstly, regardless of the values of $P$ and $M_r$, the BER upper-bounds are tight at moderate to high SNRs. Secondly, given $M_r=2$, the simulated MLD BER approaches the upper bound when $\gamma_s>12$ dB for $P=2$ and $\gamma_s>10$ dB for $P=3$. Moreover, larger $P$ and $M_r$ lead to higher diversity order and improved BER performance, thus validating our analytical diversity order results in Subsection \ref{Section4-2}. Finally, it can be observed that the case of $\{M_r,P\}=\{3,2\}$ exhibits about $2$ dB SNR gain at a BER of $10^{-4}$ over the $\{M_r,P\}=\{2,3\}$ curve, while these two curves can attain the same slope. This implies that when the system's diversity order is fixed, one can achieve higher effective coding gain by exploiting a larger number of RAs.
%%%%%%%%%%%%%%%%%%%%%%%%%%%% Figure 3 %%%%%%%%%%%%%%%%%%%%%%%%%%%%%%%
%%%%%%%%%%%%%%%%%%%%%%%%%%%%%%%%%%%%%%%%%%%%%%%%%%%%%%%%%%%%%%%%%%%%%
%%%%%%%%%%%%%%%%%%%%%%%%%%%% Figure 4 %%%%%%%%%%%%%%%%%%%%%%%%%%%%%%%
\begin{figure}[htbp]
\vspace{-1em}
\centering
\includegraphics[width=0.9\linewidth]{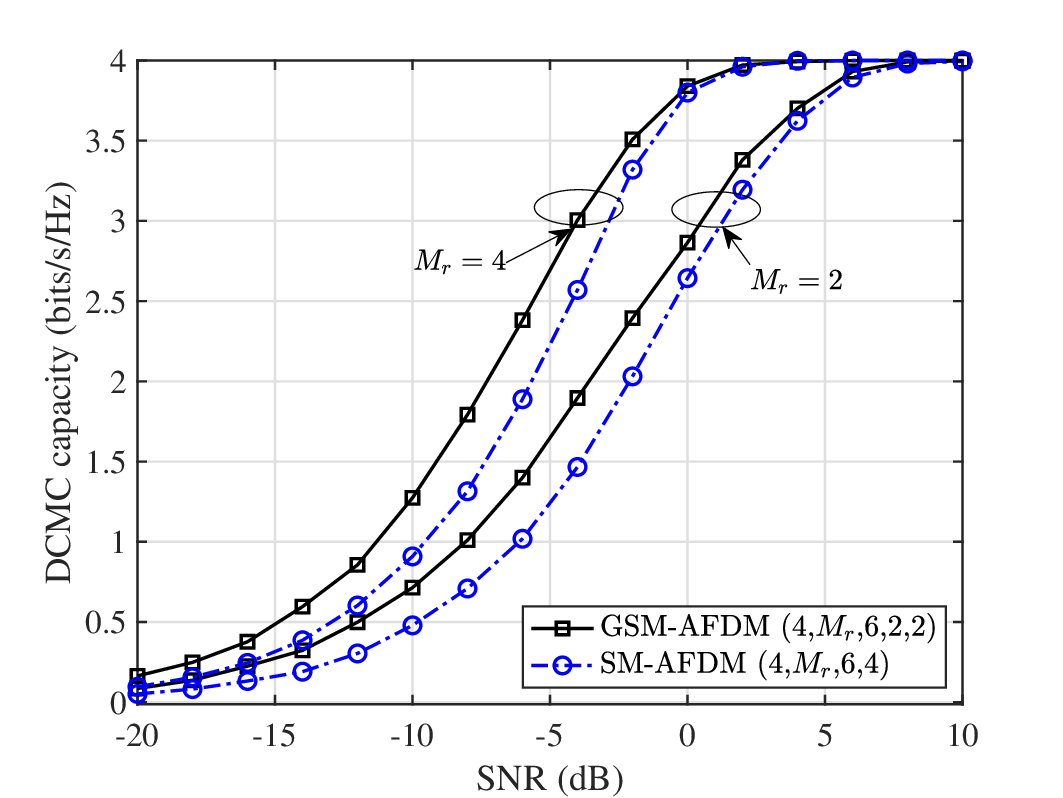}
\caption{DCMC capacity of GSM-AFDM $(4,M_r,6,2,2)$ and SM-AFDM $(4,M_r,6,4)$ with $M_r=\{2,4\}$ operating at $4$ bits/s/subcarrier.}
\label{Figure4}
%\vspace{0.5em}
\end{figure}
%%%%%%%%%%%%%%%%%%%%%%%%%%%%%%%%%%%%%%%%%%%%%%%%%%%%%%%%%%%%%%%%%%%%%

In Fig. \ref{Figure4}, the DCMC capacity of both GSM-AFDM $(4,M_r,6,2,2)$ and SM-AFDM $(4,M_r,6,4)$ associated with $M_r=\{2,4\}$ and $P=2$ are compared. It can be observed from Fig. \ref{Figure4} that the asymptotic capacity of the above schemes is $4$ bits/s/subcarrier, independent of the values of $M_r$. Moreover, given the number of RAs, it can be found that our GSM-AFDM always achieves higher DCMC capacity. This is because the constellation order of SM-AFDM is higher. Therefore, Fig. \ref{Figure4} implies that our GSM-AFDM exhibits higher coding gain than SM-AFDM systems in the case of fixed values of $PM_r$.

\begin{figure}[htbp]
\vspace{-1em}
\centering
\includegraphics[width=0.9\linewidth]{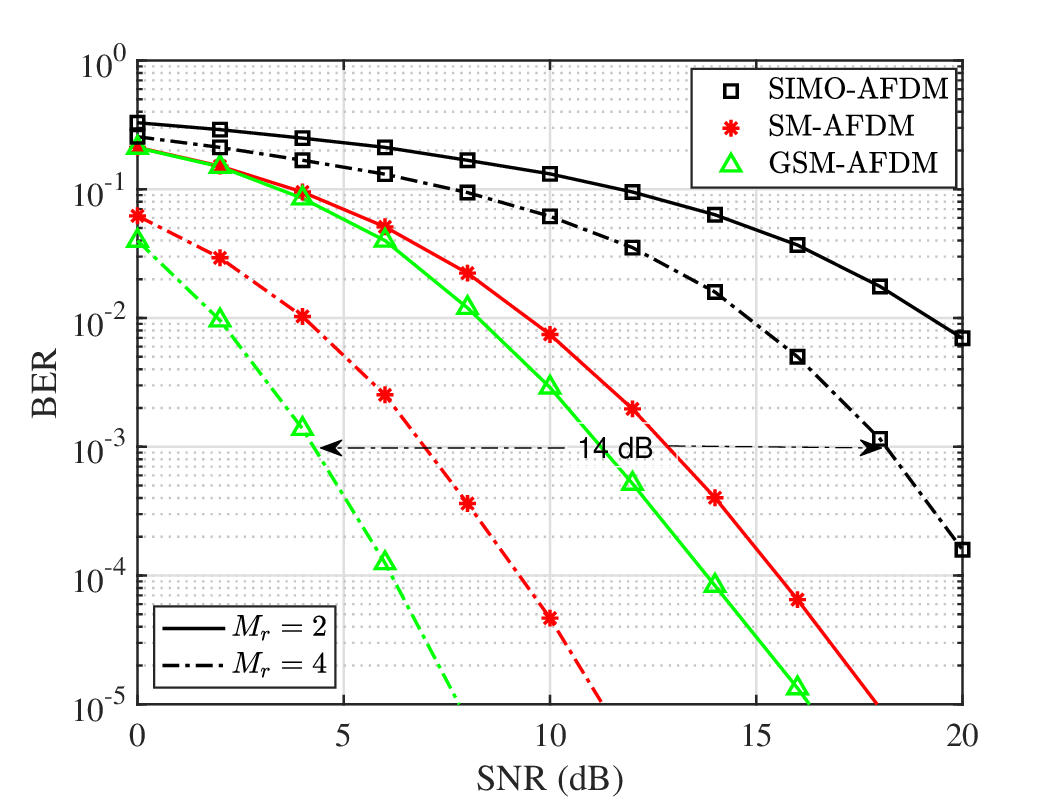}
\caption{BER performance of SIMO-AFDM $(1,M_r,6,64)$, SM-AFDM $(4,M_r,6,16)$ and GSM-AFDM $(4,M_r,6,2,4)$  with $M_r=\{2,4\}$ using MLD at the same data rate of $6$ bits/s/subcarrier.}
\label{Figure5}
\vspace{-1em}
\end{figure}

Fig. \ref{Figure5} depicts the BER performance of SIMO-AFDM $(1,M_r,6,64)$, SM-AFDM $(4,M_r,6,16)$ and GSM-AFDM $(4,M_r,6,2,4)$ with $M_r=\{2,4\}$ by employing MLD, where the corresponding transmission rate is $6$ bits/s/subcarrier. Furthermore, the maximum normalized delay shift $l_\text{max}=1$ and $P=3$ paths are considered. It can be observed that, given the modulation scheme, using more RAs can result in improved BER performance, since a higher diversity order can be achieved. Explicitly, given $M_r=4$ and BER of $10^{-3}$, the proposed GSM-AFDM attains about $2.5$ dB and $14$ dB SNR gain, respectively, while the above-mentioned BER curves have the same slope. This is because lower-order constellations are invoked in GSM-AFDM systems. Similar to the findings of Fig. \ref{Figure4}, it is demonstrated that the GSM-AFDM achieves higher coding gains compared to its conventional counterparts.

%%%%%%%%%%%%%%%%%%%%%%%%%%%% Figure 5 %%%%%%%%%%%%%%%%%%%%%%%%%%%%%%%
%%%%%%%%%%%%%%%%%%%%%%%%%%%%%%%%%%%%%%%%%%%%%%%%%%%%%%%%%%%%%%%%%%%%%
%%%%%%%%%%%%%%%%%%%%%%%%%%%% Figure 6 %%%%%%%%%%%%%%%%%%%%%%%%%%%%%%%
\begin{figure}[htbp]
\vspace{-2em}
\centering
\includegraphics[width=0.9\linewidth]{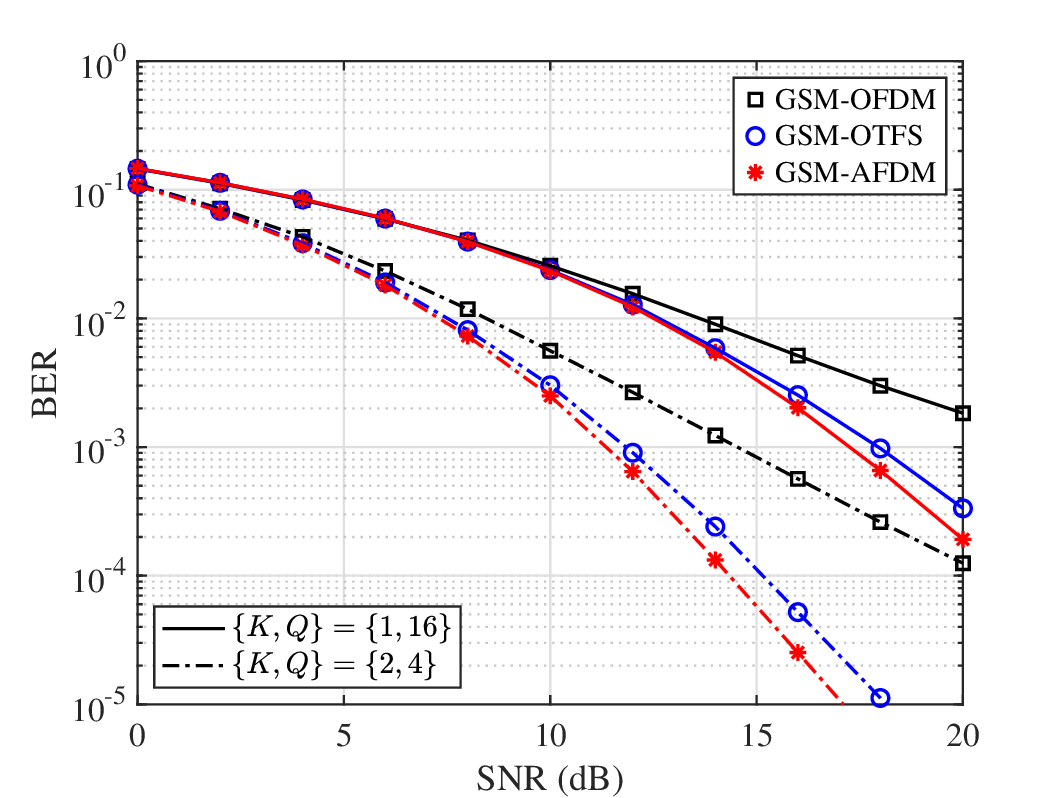}
\caption{BER performance of GSM-OFDM $(4,4,64,K,Q)$, GSM-OTFS $(4,4,8,8,K,Q)$ and GSM-AFDM $(4,4,64,K,Q)$ using LMMSE-MLD with different settings of $\{K,Q\}$ at the same rate of $6$ bits/s/subcarrier.}
\label{Figure6}
%\vspace{-1em}
\end{figure}
%%%%%%%%%%%%%%%%%%%%%%%%%%%%%%%%%%%%%%%%%%%%%%%%%%%%%%%%%%%%%%%%%%%%%

Next, we evaluate the BER performance of large-scale GSM-AFDM systems by invoking our proposed detectors, where $P=4$ paths are utilized unless specifically defined. The GSM-OTFS systems are parameterized as $(M_t,M_r,M,N,K,Q)$ and we have $M_\text{OTFS}N_\text{OTFS}=N_\text{AFDM}$. Furthermore, the bandwidths of GSM-AFDM and GSM-OTFS systems are the same, yielding $\Delta f_\text{OTFS}=16$ kHz. In Fig. \ref{Figure6}, we investigate the BER performance of GSM-OFDM $(4,4,64,K,Q)$, GSM-OTFS $(4,4,8,8,K,Q)$ and GSM-AFDM $(4,4,64,K,Q)$ employing LMMSE-MLD and different settings of $\{K,Q\}$.

From Fig. \ref{Figure6}, we have the following observations. Firstly, at a fixed combination of $\{K,Q\}$, GSM-AFDM is capable of attaining better BER performance compared to GSM-OFDM and GSM-OTFS. Specifically, with $\{K,Q\}=\{2,4\}$ at a BER of $10^{-3}$, both GSM-OTFS and GSM-AFDM can attain about $4$ dB gain compared to GSM-OFDM. This is because OFDM-based systems suffer from the ICI introduced by high-mobility channels. Moreover, in the cases of $\{K,Q\}=\{2,4\}$ and $\{K,Q\}=\{1,16\}$, GSM-AFDM can attain about $1$ dB SNR gain over the GSM-OTFS scheme at BERs of $10^{-5}$ and $3\times 10^{-4}$, respectively. This observation is due to the property that GSM-AFDM can achieve a full diversity order of $P(M_t-M_r+1)$, while the asymptotic of GSM-OTFS is only $(M_t-M_r+1)$. Finally, regardless of the modulation scheme, $\{K,Q\}=\{2,4\}$ can achieve better BER performance compared to $\{K,Q\}=\{1,16\}$, since a lower constellation order is invoked in $\{K,Q\}=\{2,4\}$ scenarios.

\begin{figure}[htbp]
\vspace{-1em}
\centering
\includegraphics[width=0.9\linewidth]{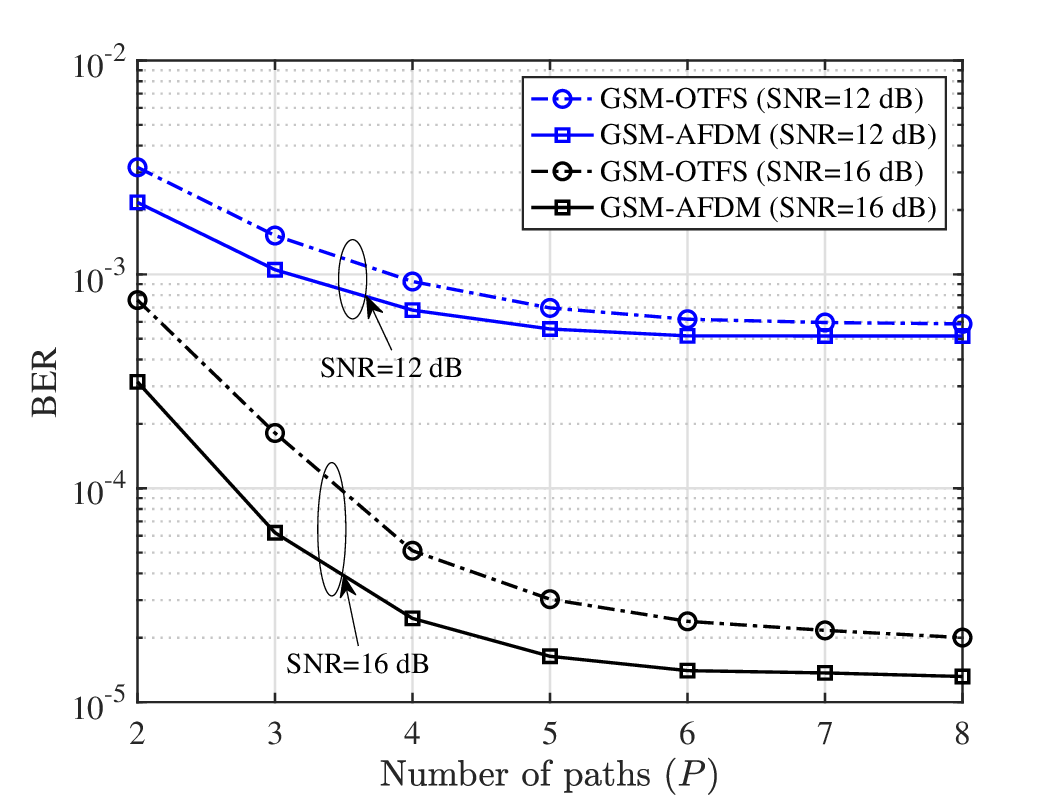}
\caption{BER performance of GSM-AFDM $(4,4,64,2,4)$ and GSM-OTFS $(4,4,8,8,2,4)$ systems versus the number of paths $P$, while LMMSE-MLD is invoked with different SNR values operating at the rate of $6$ bits/s/subcarrier.}
\label{Figure7}
%\vspace{-1em}
\end{figure}

Next, in Fig. \ref{Figure7}, we compare the BER performance of both GSM-AFDM $(4,4,64,2,4)$ and GSM-OTFS $(4,4,8,8,2,4)$ systems versus the number of paths $P$, where the LMMSE-MLD is employed. It can be observed from Fig. \ref{Figure7} that a higher SNR always leads to a BER performance gain, given the value of $P$ and the modulation order. Moreover, if the modulation scheme and the SNR are fixed, a higher value of $P$ tends to yield an improved BER performance. This is because a higher $P$ leads to a higher diversity order, which is consistent with our analytical results in Subsection \ref{Section4-2}. However, the BERs remain constant as $P$ increases since the LMMSE equalizer introduces estimation errors \cite{kay1993statistical}. Furthermore, for a given SNR, the GSM-AFDM consistently achieves better BER performance than GSM-OTFS, regardless of the value of $P$. This is because the diversity order of GSM-AFDM is $P(M_t-M_r+1)$, while the asymptotic diversity order of GSM-OTFS systems is $(M_t-M_r+1)$, which is consistent with our findings in Fig. \ref{Figure6}.
%%%%%%%%%%%%%%%%%%%%%%%%%%%% Figure 7 %%%%%%%%%%%%%%%%%%%%%%%%%%%%%%%
%%%%%%%%%%%%%%%%%%%%%%%%%%%%%%%%%%%%%%%%%%%%%%%%%%%%%%%%%%%%%%%%%%%%%

\begin{figure}[htbp]
\vspace{-1em}
\centering
\includegraphics[width=0.9\linewidth]{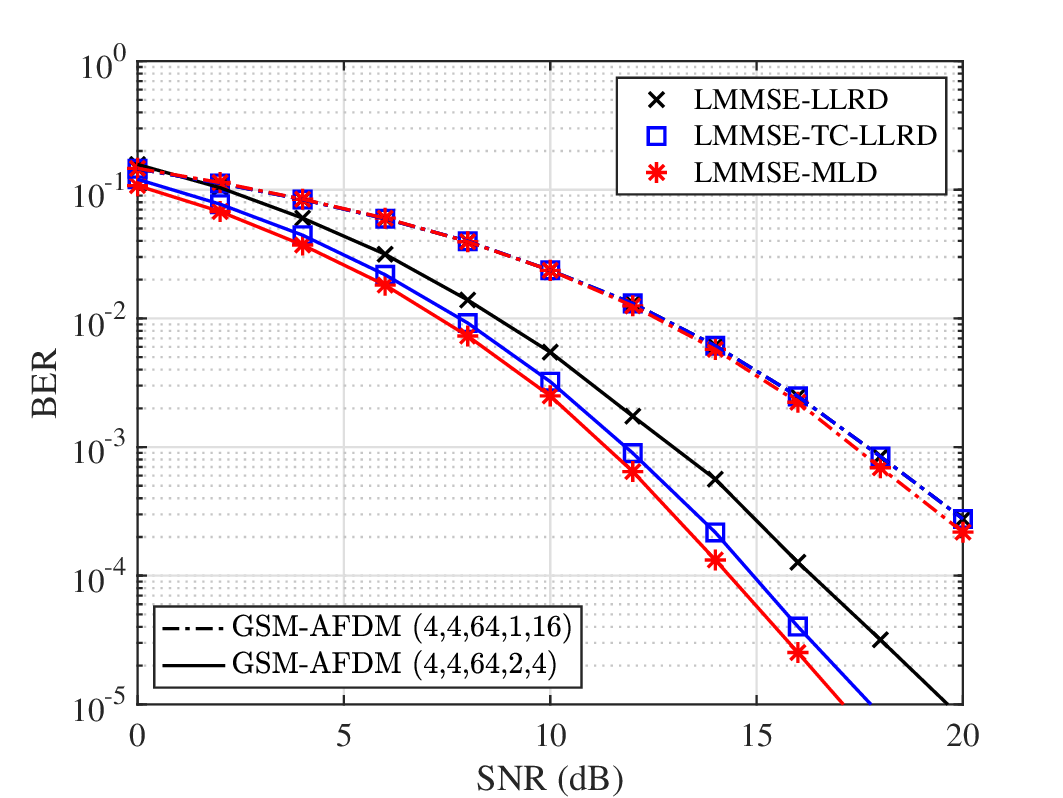}
\caption{BER performance of GSM-AFDM $(4,4,64,K,Q)$ using LMMSE-LLRD, LMMSE-TC-LLRD and LMMSE-MLD with different settings of $\{K,Q\}$ \textcolor{black}{at} the same rate of $6$ bits/s/subcarrier.}
\label{Figure8}
\vspace{-1em}
\end{figure}

Then, we compare the proposed low-complexity detectors. In Fig. \ref{Figure8}, we investigate the BER performance of LMMSE-LLRD, LMMSE-TC-LLRD and LMMSE-MLD, where we invoke GSM-AFDM $(4,4,64,K,Q)$ and different settings of $\{K,Q\}$. \textcolor{black}{The LMMSE-LLRD is used as a benchmark to evaluate the effectiveness of our proposed TC algorithm.} We have the following observations from Fig. \ref{Figure8}. Firstly, given a setting of $\{K,Q\}$, our proposed LMMSE-TC-MLD approaches the LMMSE-MLD curve, while the conventional LMMSE-LLRD yields the worst BER. Secondly, for the BER value of $10^{-5}$ along with GSM-AFDM $(4,4,64,2,4)$, the proposed LMMSE-TC-LLRD obtains about $2$ dB SNR gain compared to LMMSE-LLRD, while LMMSE-TC-LLRD only exhibits $0.5$ dB SNR loss over the LMMSE-MLD. The observations mentioned above illustrate that our proposed TC technique can avoid catastrophic TAP decisions. Finally, the BER performance of LMMSE-TC-LLRD and LMMSE-LLRD of $\{K,Q\}=\{1,16\}$ is identical. This is because we have $\binom{M_t}{K}=C$, i.e., no unused TAPs are under this scenario.
%%%%%%%%%%%%%%%%%%%%%%%%%%%% Figure 8 %%%%%%%%%%%%%%%%%%%%%%%%%%%%%%%
%%%%%%%%%%%%%%%%%%%%%%%%%%%%%%%%%%%%%%%%%%%%%%%%%%%%%%%%%%%%%%%%%%%%%
%%%%%%%%%%%%%%%%%%%%%%%%%%%% Figure 9 %%%%%%%%%%%%%%%%%%%%%%%%%%%%%%%
\begin{figure}[htbp]
\vspace{-1em}
\centering
\includegraphics[width=0.9\linewidth]{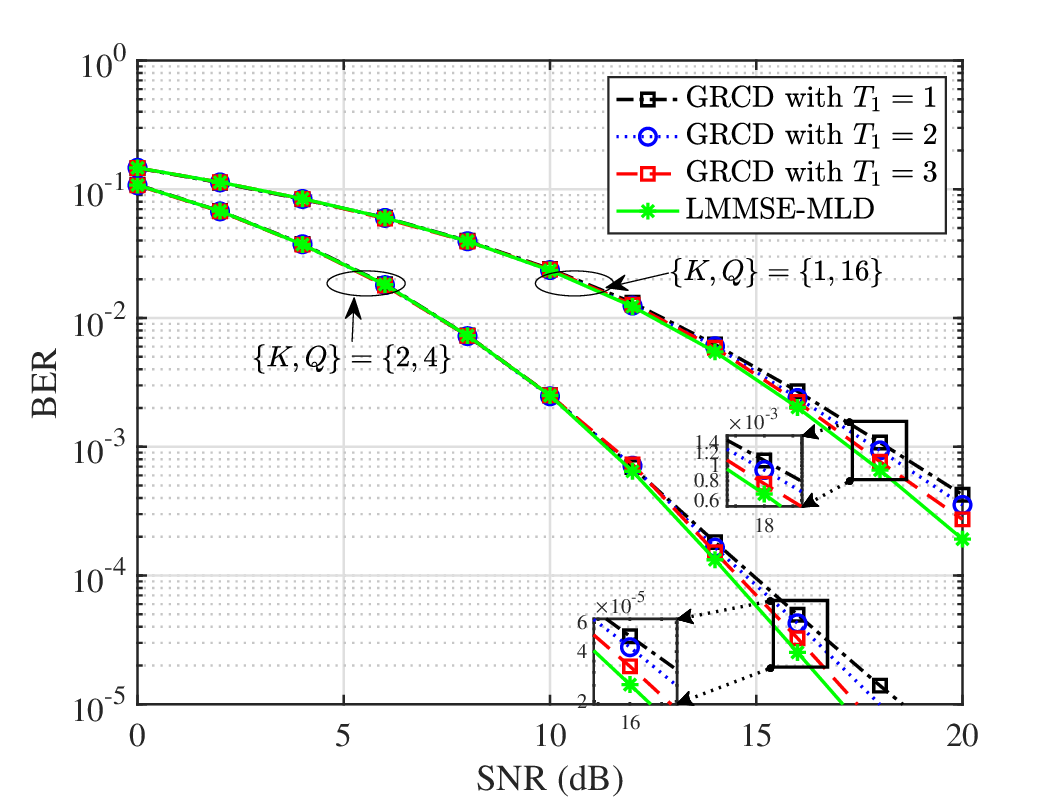}
\caption{BER performance of GSM-AFDM $(4,4,64,K,Q)$ using GRCD, while different settings of $\{K,Q\}$ and GRCD iterations $T_1$ are invoked under the rate of $6$ bits/s/subcarrier.}
\label{Figure9}
\vspace{0.5em}
\end{figure}
%%%%%%%%%%%%%%%%%%%%%%%%%%%%%%%%%%%%%%%%%%%%%%%%%%%%%%%%%%%%%%%%%%%%%

Fig. \ref{Figure9} characterizes the BER performance of the LMMSE-MLD and GRCD conceived for the GSM-AFDM $(4,4,64,K,Q)$ system. We observe from Fig. \ref{Figure9} that both the proposed GRCD associated with $T_1=1$ and $T_1=2$ are capable of attaining near-LMMSE-MLD BER performance. Moreover, regardless of the settings of $\{K,Q\}$, the BER performance of GRCD using $T_1=1$ and $T_1=3$ is very close. We emphasize that $T_1$ is the maximum number of GRCD iterations in \textbf{Algorithm  \ref{alg-grcd}}.

\begin{figure}[htbp]
\vspace{-1em}
\centering
\includegraphics[width=0.9\linewidth]{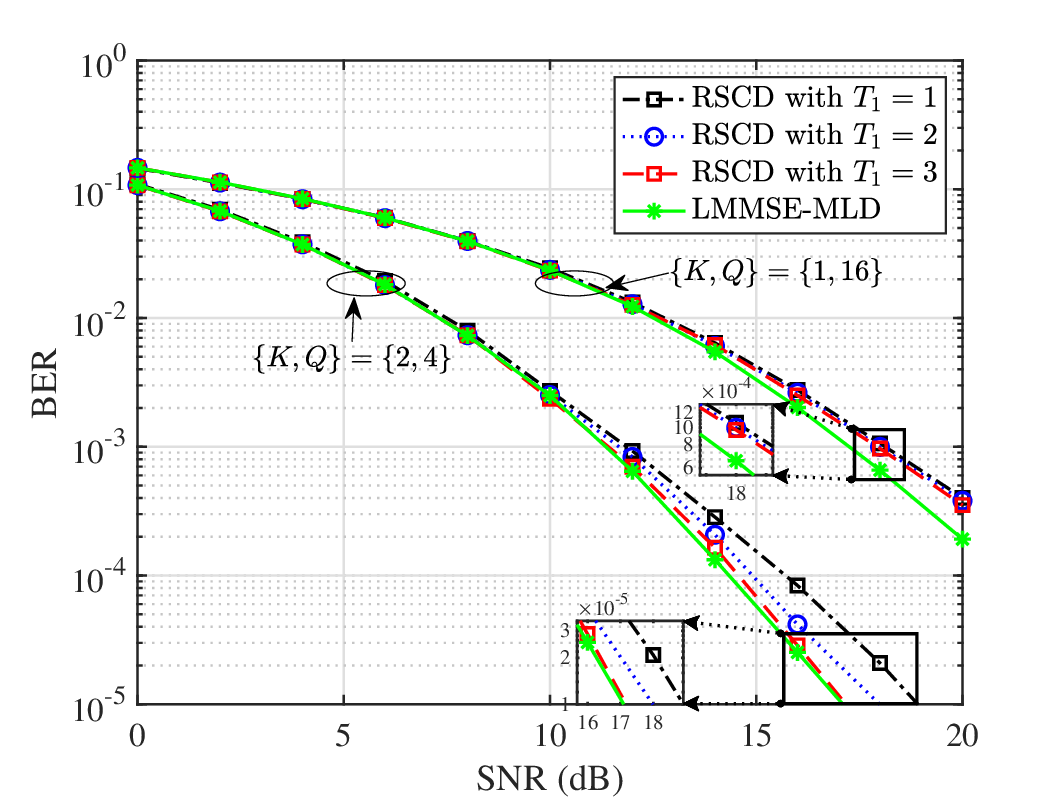}
\caption{BER performance of GSM-AFDM $(4,4,64,K,Q)$ using RSCD, while different settings of $\{K,Q\}$ and RSCD iterations $T_2$ are invoked under the rate of $6$ bits/s/subcarrier.}
\label{Figure10}
\vspace{-1em}
\end{figure}

The BER performance of the RSCD is shown in Fig. \ref{Figure10}, where LMMSE-MLD is adopted as the benchmark, while all other parameters are the same as in Fig. \ref{Figure9}. In Fig. \ref{Figure10}, we observe that higher values of $T_2$ yield improved BER performance. Specifically, when $\{K,Q\}=\{2,4\}$, RSCD using $T_2=3$ achieves gains of about $3$ dB and $1$ dB over the $T_2=1$ and $T_2=2$ scenarios, respectively. Moreover, regardless of the settings of $\{K,Q\}$, RSCD using $T_2=3$ is capable of attaining a near-LMMSE-MLD BER performance. Hence, we conclude from Fig. \ref{Figure9} and Fig. \ref{Figure10} that RSCD with $T_2=2$ and $T_2=3$ can be utilized to achieve a good BER performance. By contrast, for GRCD $T_1=1$ iteration is sufficient, as demonstrated in Fig. \ref{Figure9}, and both of them have a considerably lower complexity than the LMMSE-MLD.

%%%%%%%%%%%%%%%%%%%%%%%%%%%% Figure 10 %%%%%%%%%%%%%%%%%%%%%%%%%%%%%%%
%%%%%%%%%%%%%%%%%%%%%%%%%%%%%%%%%%%%%%%%%%%%%%%%%%%%%%%%%%%%%%%%%%%%%
%%%%%%%%%%%%%%%%%%%%%%%%%%%% Figure 11 %%%%%%%%%%%%%%%%%%%%%%%%%%%%%%%
\begin{figure}[htbp]
\centering
\includegraphics[width=0.9\linewidth]{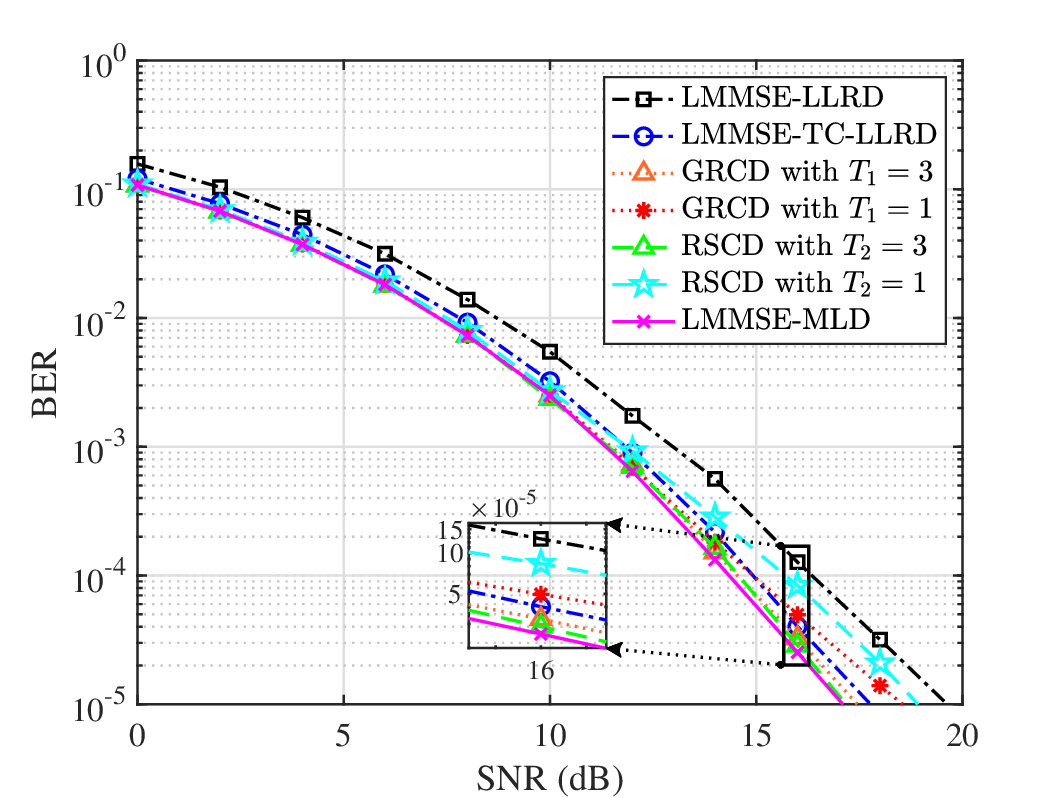}
\caption{BER comparison of GSM-AFDM $(4,4,64,2,4)$ using LMMSE-LLRD, LMMSE-TC-LLRD, GRCD with $T_1=\{1,3\}$, \textcolor{black}{RSCD with $T_2=\{1,3\}$} and LMMSE-MLD operating at the rate of $6$ bits/s/subcarrier.}
\label{Figure11}
\vspace{0em}
\end{figure}
%%%%%%%%%%%%%%%%%%%%%%%%%%%%%%%%%%%%%%%%%%%%%%%%%%%%%%%%%%%%%%%%%%%%%

\textcolor{black}{To further compare our proposed detectors, in Fig. \ref{Figure11} we evaluate the BER performance of LMMSE-LLRD, LMMSE-TC-LLRD, GRCD associated with $T_1=1$, RSCD with $T_2=\{1,3\}$ and LMMSE-MLD in GSM-AFDM $(4,4,64,2,4)$ systems, yielding a rate of $6$ bits/s/subcarrier. At the BER of $10^{-5}$, it can be observed that RSCD with $T_1=1$ performs about $0.5$ dB, $1$ dB, $1.2$ dB and $1.5$ dB worse than the GRCD with $T_1=1$, LMMSE-TC-LLRD, GRCD with $T_1=3$ and the LMMSE-MLD, respectively. Moreover, LMMSE-LLRD achieves the worst BER performance compared to its counterparts, resulting in about $3$ dB SNR loss with reference to LMMSE-MLD. This is because our LMMSE-LLRD only exploits the values of $M_t$ and $K$, but it may not be possible to avoid catastrophic TAP decisions. Furthermore, the proposed RSCD with $T_2=3$ and GRCD with $T_1=3$ achieve nearly identical BER performance  to that of LMMSE-MLD with lower complexity, and the BER performance of RSCD with $T_2=3$ is slightly lower than that of GRCD with $T_1=3$.}

\begin{figure}[htbp]
\vspace{-1em}
\centering
\includegraphics[width=0.9\linewidth]{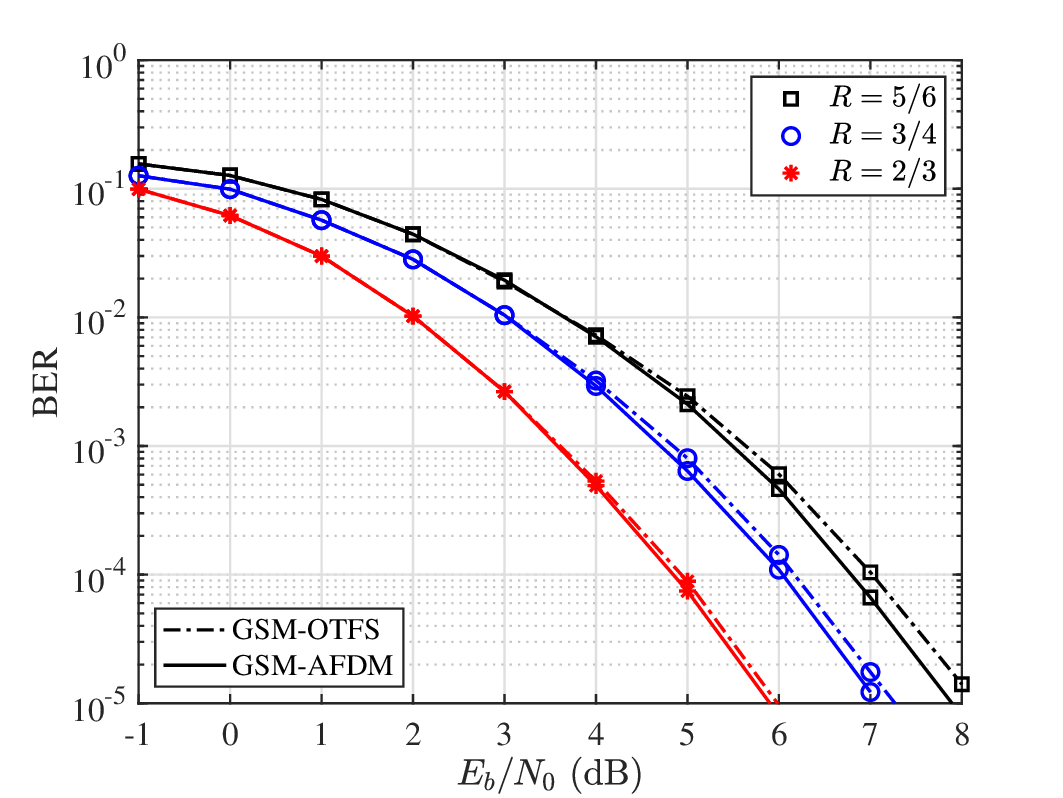}
\caption{BER performance of different rate-LDPC-coded GSM-AFDM $(4,4,128,2,4)$ and GSM-OTFS $(4,4,8,16,2,4)$ systems invoking LMMSE equalizer and soft LLR detector operating at $6$ bits/s/subcarrier.}
\label{Figure12}
\vspace{-1em}
\end{figure}

To illustrate the flexibility of the GSM-AFDM and further compare GSM-AFDM and GSM-OTFS, the BER performance of IEEE 802.11n LDPC coded \cite{6131118} GSM-AFDM $(4,4,128,2,4)$ and GSM-OTFS $(4,4,8,16,2,4)$ systems employing an LMMSE equalizer along soft LLR detector and with coding rates of $5/6$, $3/4$, and $2/3$ are characterized in Fig. \ref{Figure12}. Explicitly, we consider the LDPC codeword length to be $648$ with $T_\text{LDPC}=5$ for the belief propagation decoder iterations. Observe from Fig. \ref{Figure12} that a lower LDPC coded rate leads to a better BER performance. Moreover, since AFDM-based systems can always attain full diversity, it can be seen that the BER curves of GSM-AFDM have the same slope. Moreover, although GSM-AFDM systems consistently achieve better BER than their GSM-OTFS counterparts, the BER performance gaps between GSM-AFDM and GSM-OTFS are reduced for lower LDPC-coded rates. This implies that the diversity order of OTFS-based systems can be enhanced to some extent upon using channel coding, which is consistent with the observations from \cite{9404861}.
%%%%%%%%%%%%%%%%%%%%%%%%%%%% Figure 12 %%%%%%%%%%%%%%%%%%%%%%%%%%%%%%%
%%%%%%%%%%%%%%%%%%%%%%%%%%%%%%%%%%%%%%%%%%%%%%%%%%%%%%%%%%%%%%%%%%%%%
%%%%%%%%%%%%%%%%%%%%%%%%%%%% Figure 13 %%%%%%%%%%%%%%%%%%%%%%%%%%%%%%%
\begin{figure}[htbp]
\vspace{-1em}
\centering
\includegraphics[width=0.9\linewidth]{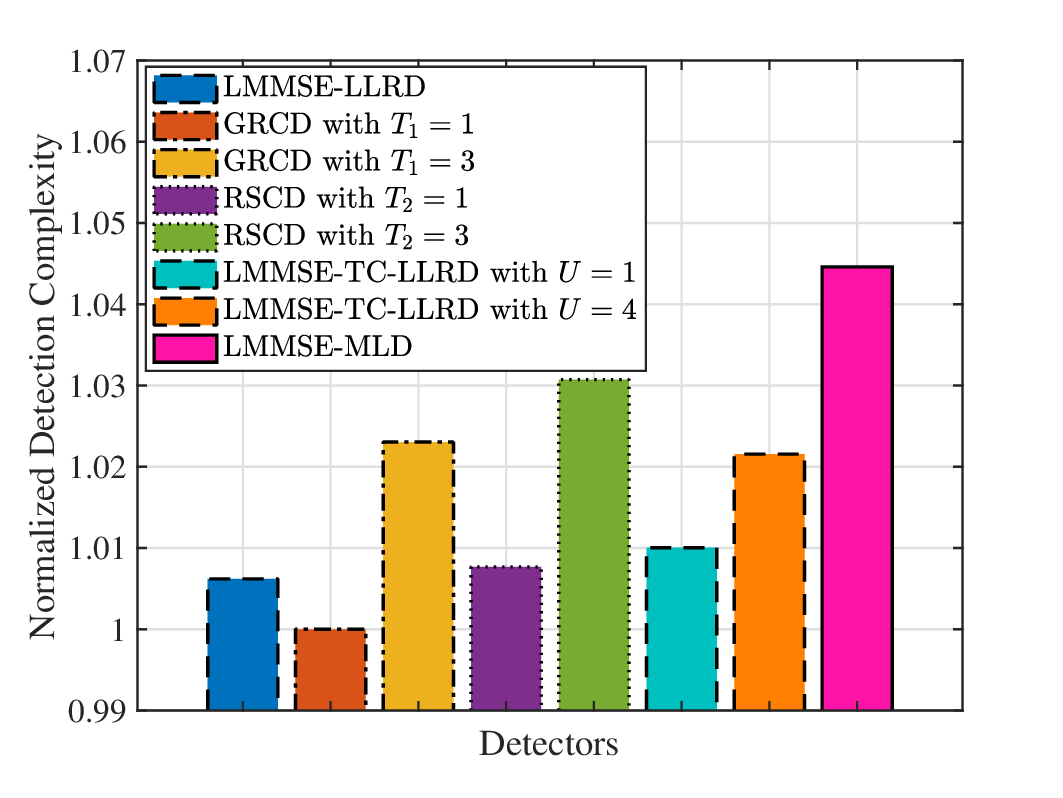}
\vspace{-2em}
\caption{\textcolor{black}{Normalized detection complexity of GSM-AFDM $(4,4,64,2,4)$ systems exploiting LMMSE-LLRD, GRCD with $T_1=\{1,3\}$, RSCD with $T_2=\{1,3\}$, LMMSE-TC-LLRD with $U=1$ (best case), LMMSE-TC-LLRD with $U=4$ (worst case) and LMMSE-MLD normalized by that of GRCD with $T_1=1$ for a transmission rate of $6$ bits/s/subcarrier.}}
\label{Figure13}
%\vspace{-1em}
\end{figure}
%%%%%%%%%%%%%%%%%%%%%%%%%%%%%%%%%%%%%%%%%%%%%%%%%%%%%%%%%%%%%%%%%%%%%

\textcolor{black}{In Fig. \ref{Figure13}, the detector complexities of LMMSE-LLRD, GRCD with $T_1=\{1,3\}$, RSCD with $T_2=\{1,3\}$, LMMSE-TC-LLRD with $U=1$ (best case), LMMSE-TC-LLRD with $U=4$ (worst case) and the LMMSE-MLD invoked in Fig. \ref{Figure11} are portrayed. We have the following observations. Firstly, the complexity of LMMSE-TC-LLRD is higher than that of its LMMSE-LLRD counterpart. This is because the proposed LMMSE-TC-LLRD includes the extra TC steps of \eqref{eq_z} and \eqref{eq_TAP}. Secondly, the complexity of GRCD with $T_1=1$ is much lower than that of its counterparts. This trend is indeed expected because our GRCD is conceived based on greedy algorithms, where each TAP is checked only once, hence avoiding repeated tests. Additionally, GRCD employs the symbol-wise detection of \eqref{eq_hard_ML}, rather than jointly estimating $K$ APM symbols, as shown in the LMMSE-MLD of \eqref{eq_MLD}. Moreover, both the best and worst cases of LMMSE-TC-LLRD and RSCD with $T_2=\{1,3\}$ have lower complexity than the LMMSE-MLD, since the TAP-indices and APM symbols are detected separately in LMMSE-TC-LLRD and RSCD. We emphasize that the actual complexity of LMMSE-TC-LLRD is between the best and worst complexities, which is also between the GRCD with $T_1=\{1,3\}$ and RSCD with $T_2=\{1,3\}$ cases. Furthermore, RSCD with $T_2=1$ can attain higher complexity compared to GRCD with $T_1=1$, since RSCD may not be able to avoid repeated tests of TAPs. For LMMSE-LLRD, its complexity is between the RSCD with $T_2=1$ and RSCD with $T_2=3$ scenarios, while it is slightly lower than RSCD with $T_2=1$ and the best case of LMMSE-TC-LLRD. Upon combining Fig. \ref{Figure11} and Fig. \ref{Figure13}, one can see that a BER vs. complexity trade-off exists among LMMSE-TC-LLRD, GRCD, RSCD and LMMSE-MLD. When we consider LMMSE-LLRD, GRCD with $T_1=3$, RSCD with $T_2=\{1,3\}$ and LMMSE-MLD, there is also a BER vs. complexity trade-off.}

%%%%%%%%%%%%%%%%%%%%%%%%%%%% Figure 14 %%%%%%%%%%%%%%%%%%%%%%%%%%%%%%%
\begin{figure}[htbp]
\vspace{-1em}
\centering
\includegraphics[width=0.9\linewidth]{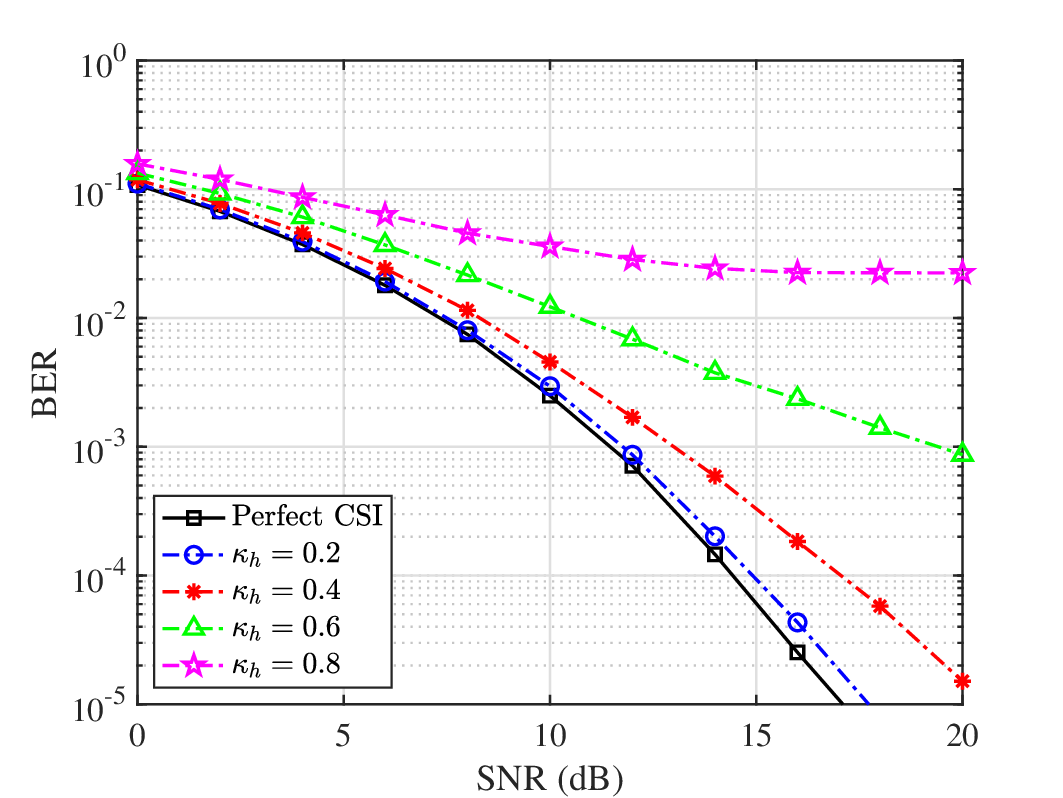}
\vspace{-1em}
\caption{\textcolor{black}{BER performance of GSM-AFDM $(4,4,64,2,4)$ using LMMSE-MLD of \eqref{eq_LLRD} with different values of the normalized channel estimation error coefficient $\kappa_h$.}}
\label{Figure14}
%\vspace{-1em}
\end{figure}
%%%%%%%%%%%%%%%%%%%%%%%%%%%%%%%%%%%%%%%%%%%%%%%%%%%%%%%%%%%%%%%%%%%%%

\textcolor{black}{We then consider the practical imperfect channel state information (CSI) scenario. Explicitly, the imperfect channel estimation can be formulated as $\hat{\pmb{h}}=\pmb{h}(1+\kappa_h\cdot\sigma_h)$ \cite{9369968}, where $0<\kappa_h<1$ denotes the normalized coefficient of channel estimation error and the complex-valued random variable $\sigma_h$ obeys a uniform distribution over the unitary circle. In Fig. \ref{Figure14}, the BER performance of GSM-AFDM $(4,4,64,2,4)$ using LMMSE-MLD under imperfect CSI cases is plotted. Specifically, the normalized channel estimation error coefficient variance is set to $\kappa_h=0.2,0.4$, $0.6$ and $0.8$, respectively. As expected, it can be observed that a higher value of $\kappa_h$ yields a worse BER performance. Explicitly, at a BER of $10^{-3}$, $\kappa_h=0.2,0.4$ and $0.6$ result in about $1$ dB, $1.5$ dB and $8.5$ dB SNR loss compared to the perfect CSI scenarios, respectively. Moreover, we observe an error floor of $2.2\times 10^{-2}$ when $\kappa_h=0.8$. This is because the BER performance is dominated by the channel estimation error at high SNRs.}

%%%%%%%%%%%%%%%%%%%%%%%%%%%% Figure 15 %%%%%%%%%%%%%%%%%%%%%%%%%%%%%%%
\begin{figure}[htbp]
\centering
\includegraphics[width=0.9\linewidth]{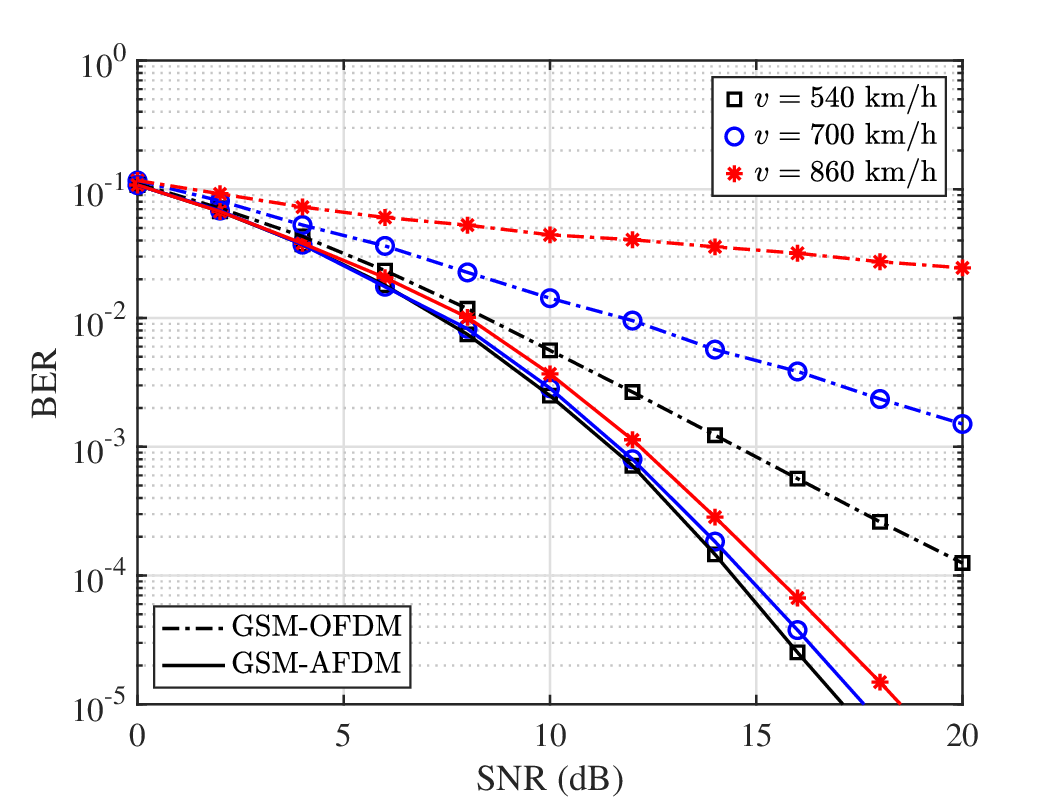}
\caption{\textcolor{black}{BER performance of GSM-AFDM and GSM-OFDM operating at $(4,4,64,2,4)$ using LMMSE-MLD under different velocities for a transmission rate of $6$ bits/s/subcarrier.}}
\label{Figure15}
%\vspace{1em}
\end{figure}
%%%%%%%%%%%%%%%%%%%%%%%%%%%%%%%%%%%%%%%%%%%%%%%%%%%%%%%%%%%%%%%%%%%%%

\textcolor{black}{In Fig. \ref{Figure15}, we present the LMMSE-MLD BER performances of GSM-AFDM and GSM-OFDM operating with system parameters of $(4,4,64,2,4)$ and velocities of $540$ km/h, $700$ km/h and $860$ km/h, respectively. From Fig. \ref{Figure15}, we have the following observations. Firstly, given the BER of $10^{-5}$, GSM-AFDM at $v=860$ km/h only suffers from about $0.5$ dB and $1$ dB SNR loss compared to the $v=700$ km/h and $v=540$ km/h scenarios, respectively. By contrast, at  BER of $2\times 10^{-3}$, we can observe that GSM-OFDM at $v=700$ km/h suffers from about $7$ dB SNR loss compared to the cases at $v=540$ km/h; the same system suffers from a BER floor of $1\times 10^{-2}$ at $v=860$ km/h. Moreover, for $v=540$ km/h, the proposed GSM-AFDM can achieve $6$ dB SNR gain compared to its GSM-OFDM counterpart at a BER of $10^{-4}$. Furthermore, we can also observe that GSM-AFDM outperforms SIMO-AFDM, SM-AFDM and GSM-OTFS at the velocity of $v=540$ km/h from Fig. \ref{Figure5} and Fig. \ref{Figure6}. Overall, the above-mentioned observations demonstrate that our proposed GSM-AFDM is robust to the ICI introduced by high-mobility channels.}
\section{Conclusions}\label{Section 6}
A GSM-AFDM transceiver was conceived, where the information bits are mapped onto both the TAP-indices and the APM symbols. We first designed the LMMSE-MLD scheme by considering large-scale GSM-AFDM systems. To alleviate the complexity of the LMMSE-MLD, the novel LMMSE-TC-LLRD, GRCD, and RSCD arrangements have been proposed, where the transmit codebook and GSM properties are exploited as the \emph{a priori} information. Explicitly, the TAP-indices and the APM symbols are detected separately, where the reliabilities of TAPs are quantified and then only a fraction of TAPs are considered. Simulation results have shown that the proposed detectors achieve near-LMMSE-MLD BER performance at a reduced complexity. Secondly, the asymptotic BER upper-bound, DCMC capacity, coding gain and diversity order of GSM-AFDM have been derived. It has been shown that the BER upper-bound is tight in the high-SNR region, and our GSM-AFDM attains full diversity. Since MIMO-AFDM can be considered as a special case of GSM-AFDM, the above analytical results can also be applied to MIMO-AFDM. Moreover, given fixed values of the diversity order and transmission rate, utilising more RAs and lower-order constellations can result in higher effective coding gains. Furthermore, the superiority of GSM-AFDM over its conventional counterparts has been validated in terms of both the BER and DCMC capacity. Finally, we have carried out a comparative study of GSM-AFDM and GSM-OTFS under both uncoded and LDPC-coded systems, showing that the proposed GSM-AFDM is capable of attaining better BER performance than its OTFS counterparts. In the future, we will investigate advanced detectors based on the sparsity of DAFT-domain channel matrices.

\renewcommand{\refname}{References}
\mbox{} 
\nocite{*}
\bibliographystyle{IEEEtran}
\bibliography{GSM_AFDM.bib}
\end{document}